\documentclass[twocolumn]{aastex62}

\usepackage[fleqn]{amsmath}
\makeatletter
\newcommand*{\rom}[1]{\expandafter\@slowromancap\romannumeral #1@}
\makeatother

\received{}
\revised{}
\accepted{}
\submitjournal{ApJ}

\shorttitle{Mira variables at maximum-light}
\shortauthors{Bhardwaj A. et al.}

\begin{document}
\title{Multiwavelength Period-Luminosity and Period-Luminosity-Color relations at maximum-light for Mira variables in the Magellanic Clouds}

\correspondingauthor{Anupam Bhardwaj}
\email{anupam.bhardwajj@gmail.com; abhardwaj@pku.edu.cn}
\author[0000-0001-6147-3360]{Anupam Bhardwaj}
\affil{Kavli Institute for Astronomy and Astrophysics, Peking University, Yi He Yuan Lu 5, Hai Dian District, Beijing 100871, China}

\author{Shashi Kanbur}
\affiliation{State University of New York, Oswego, NY 13126, USA}
\author{Shiyuan He}
\affiliation{Institute of Statistics and Big Data, Renmin University of China, Beijing 100872, China}
\author{Marina Rejkuba}
\affiliation{European Southern Observatory, Karl-Schwarzschild-Stra\ss e 2, 85748, Garching, Germany}
\author{Noriyuki Matsunaga}
\affiliation{Department of Astronomy, The University of Tokyo, Tokyo 113-0033, Japan}
\author{Richard de Grijs}
\affiliation{Department of Physics and Astronomy, Macquarie University, Balaclava Road, Sydney, NSW 2109, Australia}
\affiliation{Research Centre for Astronomy, Astrophysics and Astrophotonics, Macquarie University, Balaclava Road, Sydney, NSW 2109, Australia}
\affiliation{International Space Science Institute -- Beijing, 1 Nanertiao, Zhongguancun, Hai Dian District, Beijing 100190, China}
\author{Kaushal Sharma}
\affiliation{Inter-University Centre for Astronomy and Astrophysics, Pune, Maharashtra 411007, India}
\author{Harinder P. Singh}
\affiliation{Department of Physics and Astrophysics, University of Delhi,  Delhi-110007, India }
\author{Tapas Baug}
\affiliation{Kavli Institute for Astronomy and Astrophysics, Peking University, Yi He Yuan Lu 5, Hai Dian District, Beijing 100871, China}
\author{Chow-Choong Ngeow}
\affiliation{Graduate Institute of Astronomy, National Central University, Jhongli 32001, Taiwan}
\author{Jia-Yu Ou}
\affiliation{Graduate Institute of Astronomy, National Central University, Jhongli 32001, Taiwan}

\begin{abstract} 
We present Period-Luminosity and Period-Luminosity-Color relations at maximum-light for Mira variables in the Magellanic Clouds using time-series data from the 
Optical Gravitational Lensing Experiment (OGLE-III) and {\it Gaia} data release 2. The maximum-light relations exhibit a scatter typically up to $\sim 30\%$ smaller 
than their mean-light counterparts. The apparent magnitudes of Oxygen-rich Miras at maximum-light display significantly smaller cycle-to-cycle variations than 
at minimum-light. High-precision photometric data for Kepler Mira candidates also exhibit stable magnitude variations at the brightest epochs while their multi-epoch 
spectra display strong Balmer emission lines and weak molecular absorption at maximum-light. The stability of maximum-light magnitudes for Miras possibly occurs due 
to the decrease in the sensitivity to molecular bands at their warmest phase. At near-infrared wavelengths, the Period-Luminosity relations of Miras display similar dispersion at 
mean and maximum-light with limited time-series data in the Magellanic Clouds. A kink in the Oxygen-rich Mira Period-Luminosity relations is found at 300 days in the $VI$-bands 
which shifts to longer-periods ($\sim 350$~days) at near-infrared wavelengths. Oxygen-rich Mira Period-Luminosity relations at maximum-light provide a relative distance modulus, 
$\Delta \mu = 0.48\pm0.08$~mag, between the Magellanic Clouds with a smaller statistical uncertainty than the mean-light relations.
The maximum-light properties of Miras can be very useful for stellar atmosphere modeling and distance scale studies provided their stability and the universality can be established
in other stellar environments in the era of extremely large telescopes.\\
\end{abstract}

\section{Introduction}

Mira variables are large-amplitude asymptotic giant branch (AGB) stars which are evolved from low- to intermediate-mass ($\sim 0.8 < M/M_\sun < 8$) stars, and trace
intermediate-age to old stellar populations in their late stage of evolution \citep{whitelock2012}. Therefore, Mira-like variables are found in all types of galaxies 
and can be easily identified thanks to their large-amplitudes ($\Delta V \gtrsim 2.5$~ mag, $\Delta I \gtrsim 0.8$~mag,  $\Delta K \gtrsim 0.4$ mag, and  $\Delta[3.6] \gtrsim 0.2$~mag) 
and typical long-periods \citep[$\sim 100-1000$~days,][]{oglelmcmira2009, riebel2010}. Miras are also separated into two classes: Oxygen-rich (O-rich) 
or Carbon-rich (C-rich) based on their surface chemistry or photometric light-curve variations. O-rich Miras exhibit a Period-Luminosity 
relation \citep[PLR,][]{glass1981} in the Large Magellanic Cloud (LMC) with an intrinsic scatter comparable to the classical Cepheids at 
near-infrared (NIR) wavelengths \citep{whitelock2008, yuan2017b, yuan2018}. At present, the calibration of the extragalactic distance scale 
using Cepheid variables provides a value of the Hubble constant \citep{riess2016, riess2019} that is in tension with the {\it Planck} results based 
on the cosmic microwave background \citep{planck2018}. NIR PLRs of Miras have been successfully employed to estimate distances to galaxies within and
beyond the Local Group \citep{rejkuba2004, menzies2015, huang2018}. Therefore, Mira variables have the potential to provide a distance scale 
independent of classical Cepheids with the absolute calibration of their PLRs in the Milky Way or the LMC.

Several earlier Mira-based studies used a dozen long-period variables (LPVs) discovered at optical wavelengths \citep{lloyd1971} to 
derive Mira PLRs and Period-Luminosity-Color (PLC) relations in the LMC at NIR wavelengths \citep{glass1981, glass1982, feast1989}. 
In the past two decades, time-domain wide-field variability surveys have revolutionized LPV studies in the Galaxy, LMC and the Small Magellanic Cloud 
\citep[SMC,][]{wood1999, wood2000, oglelmcmira2009, oglesmcmira2011}. The LPVs were found to exhibit five distinct, almost-parallel sequences in the PL plane in both Magellanic Clouds 
\citep{wood1999, ita2004a, ita2004b, fraser2008, oglelmcmira2009, soszy2013a}, and in the Milky Way bulge \citep{soszy2013}. 
LPV sequences consist of Miras, semi-regular variables, OGLE small-amplitude red giants, and long secondary period variables \citep[for details, see ][]{soszynski2007, trabucchi2019}.
Miras are the brightest and largest amplitude LPVs that pulsate in the fundamental mode and come in two major groups, separated by the dominant chemistry in the envelope.
Multiwavelength data for Miras in the LMC have been used in a wide range of studies: examples include an investigation of the complex effects of
circumstellar extinction \citep{ita2011}, the derivation and application of NIR PLRs for O-rich Miras with scatter as low as $0.12$~mag in the $K_s$-band 
for distance scale applications \citep{yuan2017b}, and the classification of Miras into O- and C-rich AGB subclasses \citep{lebzelter2018}.  

In the theoretical framework, one-dimensional linear, non-adiabatic, radial pulsation models have allowed a significant progress to understanding 
and identifying LPV sequences \citep{wood2015, trabucchi2017}. 
While these models are able to reproduce periods for early AGB stars pulsating in overtone modes, the properties of fundamental mode Miras are strongly affected 
by the internal structure of the models \citep{trabucchi2019}. On the other hand, non-linear fundamental-mode pulsation codes that are based on self-excited pulsation models and opacity-sampling 
treatment of radiation transport, can simulate fundamental-mode Mira-type pulsations and fit observed light curves and amplitudes \citep{ireland2011}. 
Recent, global three-dimensional radiation-hydrodynamical simulations show that convection, waves, and shocks all contribute to increasing the stellar 
radius and to levitating material into layers where dust can form \citep{bladh2015, freytag2017, liljegren2018}. These models self-consistently describe 
convection, fundamental-mode radial pulsations, and atmospheric shocks, and show that the pulsation periods are in good agreement with the observed Mira 
PLRs \citep{freytag2017}. 

The spectral features of Miras as shown for prototype {\it o Ceti} \citep{joy1926}, include hydrogen emission lines with the greatest intensity at maximum-light 
(henceforth max-light) and the weakest around minimum-light \citep[henceforth min-light,][]{joy1954, castelaz1997, yao2017}. In the case of O-rich Miras, 
the Balmer emission-line flux of $H\alpha$ is the weakest among the four 
Balmer lines while $H\delta$ has the greatest intensity at max-light. In contrast, the titanium oxide (TiO) molecular bands are typically weaker at max-light 
\citep{castelaz1997, castelaz2000}. The Balmer line increment in O-rich Mira variables was initially attributed to TiO absorption 
\citep{merrill1940, joy1947, gillet1988}. Later, hydrodynamical models suggested that radiative transfer effects can cause this increment 
when the hydrogen lines are formed in a shocked atmosphere \citep{bowen1988, luttermoser1992, richter2003}. Spectro-interferometric studies of O-rich Miras
suggest that the photospheric radius and the effect of unstable water vapors increase from the max- to min-light \citep{wittkowski2018}. 
Therefore, phase-dependent spectroscopic and photometric studies of Mira variables can be used to probe both stellar atmosphere variations and pulsation mechanisms.

Using photometric data, \citet{kanbur1997} carried out the only study of PL and PLC relations as a function of phase for LMC Miras in the $JHK_s$ bands. They 
showed that the dispersion in these relations is smaller than that of their counterparts at mean-light. The Mira sample used by \citet{kanbur1997} was limited 
to 48 stars with sparsely sampled NIR light curves. Therefore, it is essential to examine the phase-dependent variations in the Mira PL and PLC relations with 
modern multiwavelength data and relate the pulsation mechanism effects to stellar atmospheric variations. For the distance measurements, infrared PLRs of Miras 
are preferred because of lower extinction, lower metallicity effects, and smaller temperature fluctuations at longer wavelengths. For example, AGB pulsation 
properties are found to be similar in metal-poor ([Fe/H]$\gtrsim-1.85$~dex) dwarf galaxies and metal-rich ([Fe/H]$\sim-0.38$~dex) LMC in the 3.6$\mu$m and 
4.5$\mu$m~bands \citep{goldman2019}. However, the wealth of data from ongoing time-domain surveys provides an opportunity to better characterize the PL 
and PLC relations for Miras also at optical wavelengths. This work aims to use the plethora of multi-wavelength data for LPVs in the Magellanic Clouds 
to explore the stability and characteristics of Mira PL and PLC relations at max-light.

The paper is organized as follows. In Section~\ref{sec:data}, we describe the data used in our analysis and provide 
a brief overview of the methodology. Our comparisons of PL and PLC relations at mean and max-light are discussed in Section~\ref{sec:mira_opt}. The stability
of max-light properties with both photometric and spectroscopic data is discussed in Section~\ref{sec:mira_max}. NIR properties of Miras at max-light and
their possible application for distance measurements are presented in Section~\ref{sec:mira_nir}.  The results are summarized in Section~\ref{sec:discuss}.

\section{Data and Methods} \label{sec:data}

\begin{figure}
\epsscale{1.2}
\plotone{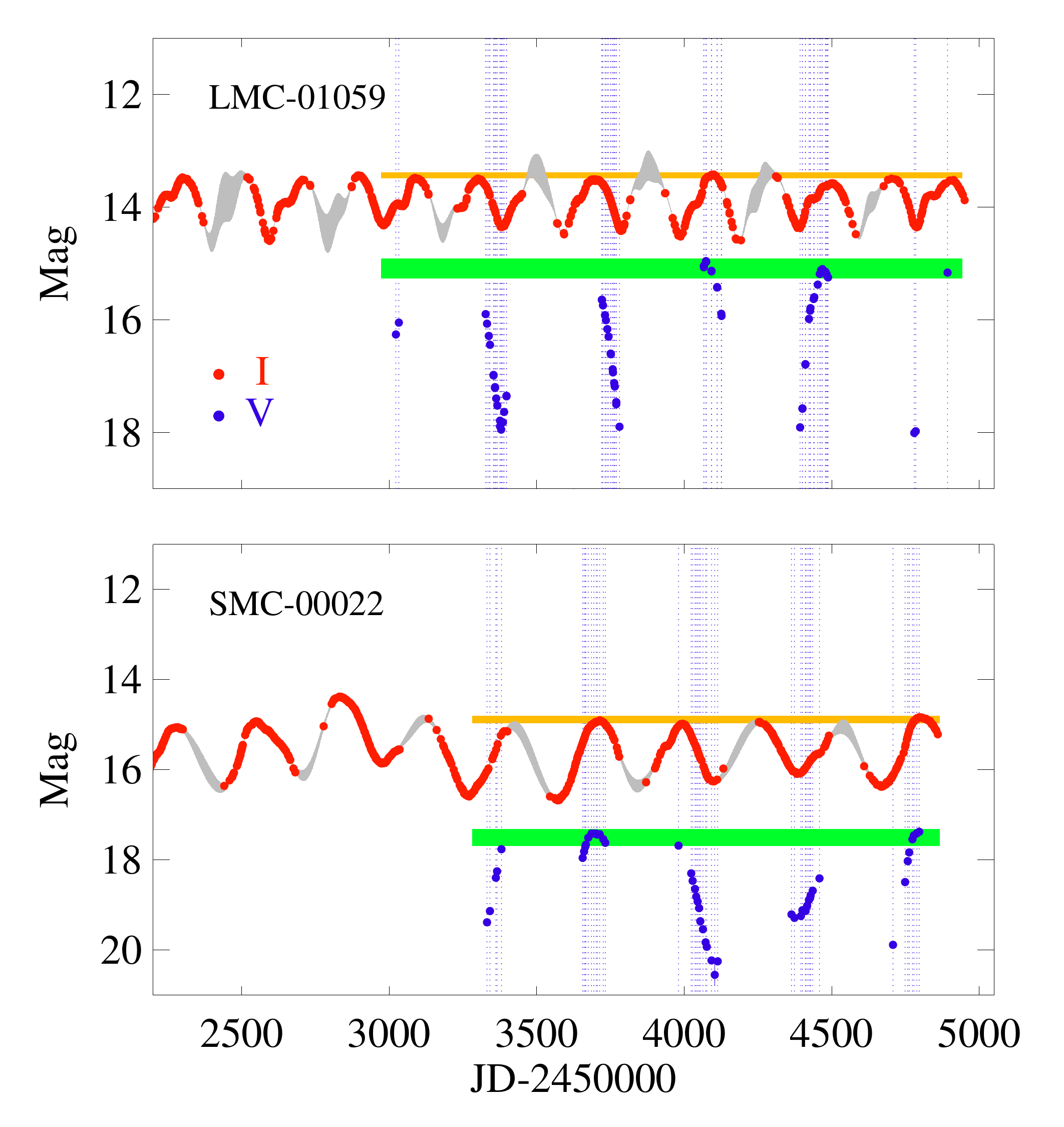}
\caption{Examples of Gaussian process regression model fitting to observed $I$-band light curves of Miras in the Magellanic Clouds. The model light curve (grey line) is used to
estimate the $I$-band magnitude corresponding to each $V$-band observation. Vertical dotted lines show the $V$ and $I$-band magnitude at the same epoch. Horizontal shaded 
lines represent the ten-percentile range used to estimate max-light magnitudes in $V$ and $I$-bands.} 
\label{gpr_lc}
\end{figure}

Time-series data for Mira variables from OGLE-\rom{3} \citep{oglelmcmira2009, oglesmcmira2011} are adopted as the primary catalog. Out of 1663/352 Miras in the LMC/SMC, 
there are 1438/302 (457/36 O-rich, 981/266 C-rich) stars with both $V$ and $I$-band light curves. On average, more than 600 $I$-band and 60 $V$-band observations are 
available for each Mira variable, which span more than a decade.
Multi-band photometry is obtained by cross-matching OGLE data with the {\it Gaia} data release 2 \citep{gaia2018, mowlavi2018} within a $2''$ search radius. 
The light curves of 650/111 Miras in the LMC/SMC are obtained with only $\sim 80$ median number of observations per Mira combined in the three {\it Gaia} filters. 
A limited amount of NIR data available for 690 Miras in the LMC \citep{yuan2017b}, and 90 Miras in the SMC \citep{ita2018} are also used in this analysis. 
The periods and O/C-rich Mira classifications are adopted from the OGLE-\rom{3} catalog. 

\begin{deluxetable*}{rrrrrrrrrrrrrrrr}
\tablecaption{Mira properties in the Magellanic Clouds using OGLE-III data. \label{tbl:mc_mag}}
\tabletypesize{\footnotesize}
\tablewidth{0pt}
\tablehead{\colhead{OGLE ID} & \colhead{Period} & \colhead{Cl}& \multicolumn{4}{c}{Max-light magnitudes}  & \multicolumn{4}{c}{Mean-light magnitudes} & \multicolumn{4}{c}{Min-light magnitudes}  & \colhead{$\Delta_{I}$}\\
 & [days] & 	& $V_M$ & $\sigma_{V_M}$ & $I_M$& $\sigma_{I_M}$&  $V$ & $\sigma_V$ & $I$ & $\sigma_I$ & $V_N$ & $\sigma_{V_N}$ & $I_N$ & $\sigma_{I_N}$& mag }
\startdata
LMC-00055&     290.90& C&     18.401&      0.094&     15.212&      0.064&     20.114&      0.088&     16.335&      0.061&     22.084&      0.092&     17.562&      0.081&      1.635\\
LMC-00082&     164.84& O&     15.869&      0.078&     13.854&      0.051&     16.423&      0.050&     14.153&      0.051&     18.394&      0.236&     14.795&      0.050&      0.906\\
LMC-00094&     332.30& C&     18.428&      0.063&     14.717&      0.053&     19.026&      0.056&     14.991&      0.056&     20.569&      0.074&     15.784&      0.057&      0.896\\
LMC-00096&     374.60& C&     17.082&      0.060&     14.086&      0.054&     17.742&      0.051&     14.519&      0.056&     19.123&      0.058&     15.225&      0.055&      1.584\\
LMC-00098&     323.10& C&     17.014&      0.071&     14.292&      0.061&     18.648&      0.054&     15.390&      0.059&     19.706&      0.069&     16.181&      0.063&      1.502\\
\enddata
\tablecomments{IDs, periods and classification (Cl: O or C-rich) are from the OGLE catalogs \citep{oglelmcmira2009, oglesmcmira2011}. Only the first five lines of the Table are presented here; 
the rest can be found in the online supplemental material.}
\end{deluxetable*}

To estimate max-light properties, we restrict our sample to light-curves with a minimum of 15 observations in a given
filter. The semi-parametric Gaussian process regression (GPR) model of \citet{he2016} is used to fit the $I$-band light curves
of OGLE-\rom{3} variables. GPR is used to model both periodic variations and stochastic signals in Mira light curves, simultaneously. 
The stochastic signals represent changes in the average light curve variations that may occur, for example, due to the formation and destruction of molecules in 
the extended atmospheres of Miras. Therefore, the GPR model can fit long term trends in Mira light curves and provide accurate periods even for sparsely sampled 
light curves \citep[see][for details]{he2016}.
These GPR fits are used as templates to predict the $I$-band magnitude corresponding to each $V$-band observation. 
Fig.~\ref{gpr_lc} displays model-fits to $I$-band light curves for
two Miras in the Magellanic Clouds. This model-fitting approach allows us to robustly estimate colors at each epoch when both $VI$-measurements are available. 

The max-light magnitudes are estimated by taking the mean of all the epochs within the brightest ten percentile of the peak-to-peak amplitude for each light curve in
each band. The standard deviation of this average is adopted as its associated uncertainty. This definition is adopted firstly because there are only a limited number 
of epochs in most bands except the $I$-band. Therefore, our light curves do not necessarily cover the true maximum, and an average of the brightest epochs along the 
light curve allows a reasonable estimate of apparent magnitude at max-light. Secondly, if the light curves are not well-sampled around max-light, the adopted error in 
the max-light magnitude will be larger, thus giving it lower weight in the PLR fits. Finally, the ten percentile variation around max-light should also compensate for the 
unaccounted phase lag in max-light between $V$- and $I$-band. Magnitudes with a subscript ($I_M$) denote max-light magnitudes throughout this paper. 
The sensitivity of the results to the adopted definition is also discussed later. Similarly, the mean-magnitudes are adopted from the GPR fits after excluding 
data points that are predicted to have large uncertainties ($>$0.1~mag). If the GPR fits are not available, mean-magnitudes are simply the average of all 
epochs after excluding extreme outliers, i.e., data points beyond 5 and 95 percentile of the light curve. The basic properties and estimated magnitudes for Miras
in the Magellanic Clouds are listed in Table~\ref{tbl:mc_mag}.

The following relations are fitted to the data:

\begin{itemize}

\item{Single-slope model : ${m} = a + b(\log P-\log P_b) + c(m_1-m_2)$ }
\item{Two-slope model : ${m} = a + b(\log P-\log P_b)_{P<P_b} + b_l(\log P-\log P_b)_{P\ge P_b} + c(m_1-m_2)$ }
\item{Quadratic model : ${m} = a + b(\log P-\log P_b) + b_l(\log P-\log P_b)^2 + c(m_1-m_2)$}.

\end{itemize}

\begin{deluxetable*}{llrrrrrrrr}
\tablecaption{The PL and PLC relations for Miras in the Magellanic Clouds using the OGLE-III data. \label{tbl:ogl_mc}}
\tabletypesize{\footnotesize}
\tablewidth{0pt}
\tablehead{\colhead{MC} & \colhead{Band} & \colhead{O/C}& \colhead{$a$} & \colhead{$b$} & \colhead{$b_l$} & \colhead{$c$} & \colhead{$\sigma$} & \colhead{$N_i$}  & \colhead{$N_f$}}
\startdata
\multicolumn{10}{c}{{Quadratic regression}}\\
\cline{1-10}
LMC               & $I$& O &      14.06$\pm$0.03      &	$-$2.13$\pm$0.10      &	$-$1.85$\pm$0.38      &                 ---      &       0.37&          454&          440\\
                & $I_M$& O &      13.36$\pm$0.02      &	$-$2.87$\pm$0.08      &	$-$1.97$\pm$0.29      &                 ---      &       0.29&          454&          441\\
            & $I,(V-I)$& O &      11.79$\pm$0.08      &	$-$4.32$\pm$0.09      &	$-$1.22$\pm$0.24      &       0.59$\pm$0.02      &       0.23&          454&          438\\
      & $I_M,(V_M-I_M)$& O &      11.72$\pm$0.06      &	$-$4.51$\pm$0.08      &	$-$2.61$\pm$0.19      &       0.62$\pm$0.02      &       0.17&          454&          439\\
\cline{1-10}
\multicolumn{10}{c}{Linear regression}\\
\cline{1-10}
LMC               & $I$& O &      14.15$\pm$0.03      &	$-$1.17$\pm$0.16      &	$-$3.36$\pm$0.22      &                 ---      &       0.36&          454&          439\\
                & $I_M$& O &      13.46$\pm$0.03      &	$-$1.87$\pm$0.12      &	$-$4.48$\pm$0.17      &                 ---      &       0.26&          454&          434\\
            & $I,(V-I)$& O &      11.91$\pm$0.08      &	$-$3.64$\pm$0.14      &	$-$4.92$\pm$0.14      &       0.58$\pm$0.02      &       0.22&          454&          439\\
      & $I_M,(V_M-I_M)$& O &      11.85$\pm$0.06      &	$-$3.30$\pm$0.10      &	$-$5.75$\pm$0.13      &       0.61$\pm$0.02      &       0.17&          454&          439\\
            & $I,(V-I)$& C &      11.15$\pm$0.13      &	$-$2.53$\pm$0.31      &                 ---      &       1.27$\pm$0.04      &       0.63&          957&          855\\
      & $I_M,(V_M-I_M)$& C &      10.59$\pm$0.09      &	$-$4.29$\pm$0.23      &                 ---      &       1.41$\pm$0.03      &       0.42&          957&          870\\
\cline{1-10}
SMC          & $I$& O &      13.93$\pm$0.09      &	$-$3.48$\pm$0.40      &                 ---      &                 ---      &       0.45&           33&           31\\
           & $I_M$& O &      13.29$\pm$0.07      &	$-$4.23$\pm$0.32      &                 ---      &                 ---      &       0.39&           33&           32\\
       & $I,(V-I)$& O &      11.50$\pm$0.25      &	$-$5.45$\pm$0.36      &                 ---      &       0.79$\pm$0.08      &       0.33&           33&           32\\
& $I_M,(V_M- I_M)$&O &      11.57$\pm$0.20      &	$-$5.24$\pm$0.29      &                 ---      &       0.86$\pm$0.09      &       0.28&           33&           32\\
       & $I,(V-I)$& C &      11.45$\pm$0.25      &	$-$1.03$\pm$0.46      &                 ---      &       1.31$\pm$0.08      &       0.53&          262&          233\\
& $I_M,(V_M- I_M)$&C &      11.07$\pm$0.16      &	$-$2.29$\pm$0.33      &                 ---      &       1.38$\pm$0.06      &       0.38&          262&          233\\
\cline{1-10}
\enddata
\tablecomments{The coefficients ($a, b, b_l, c$) are defined in Section~\ref{sec:data}. $\sigma$: dispersion (mag). $N_i$: initial number of sources. $N_f$: final number after excluding outliers.}
\end{deluxetable*}

\noindent Here $m$ is the apparent magnitude and $m~=~m_1~\textrm{or}~m_2$ in the case of two bands. The break-period ($P_b$=300 days) represents the adopted zero-point and will be determined 
using statistical methods in the next section. In the case of PLRs, a color term ($c=0$) is not included in the above equations. An iterative outlier removal method is 
adopted throughout this study: this excludes the star with the largest residual from the fit until all residuals are within $3\sigma$ dispersion across the fitted model. 

Extinction corrections are applied using the reddening maps of \citet{hasch2011} which give the color excess $E(V-I)$ for Miras in the Magellanic 
Clouds. The median value of $E(V-I)$ is taken if these maps do not provide the color-excess value within a $2''$ search radius for each Mira. For a given color excess, the extinction 
in different filters is estimated by adopting the \citet{card1989} extinction-law which gives a total-to-selective absorption ratio 
($R_{\lambda}$) of $R_{V,I,J,H,K}=\ 2.40,\ 1.41,\ 0.69,\ 0.43,\ 0.28$, at different wavelengths. The foreground reddening towards the LMC is small, specially at longer wavelengths,
and it is expected to be significantly smaller than circumstellar reddening associated with Mira variables \citep{ita2011}.\\

\section{Miras at maximum light} \label{sec:mira_opt}

The spectral type of classical Cepheid variables is independent of their pulsation period at max-light \citep{code1947}. Following the work of \citet{smk1993} on Cepheid 
properties at max-light and assuming Mira pulsations, like Cepheids, are envelope phenomena, \citet{kanbur1997} speculated that the scatter in the PLRs at max-light arises 
predominantly from the range of total masses of Miras whereas both the range of core masses and total masses contribute to the scatter at mean-light. 
However, Miras exhibit large-amplitude pulsations, temporal variations in their stellar diameters, and extended dynamical atmospheres with complex molecular 
layers \citep[see the review by][]{hofner2018}. Pulsations in Mira-like variables are also strongly coupled with poorly understood processes such as dust formation and mass loss. 
Nevertheless, while investigating max-light Mira PLRs for the first-time, \citet{kanbur1997} found that the $J$-band PLR exhibits significantly 
smaller scatter at max-light than at mean-light for Miras in the LMC. Here, we use optical data for Miras for the first time to examine this in the following subsections.

\begin{figure*}
\epsscale{1.2}
\plotone{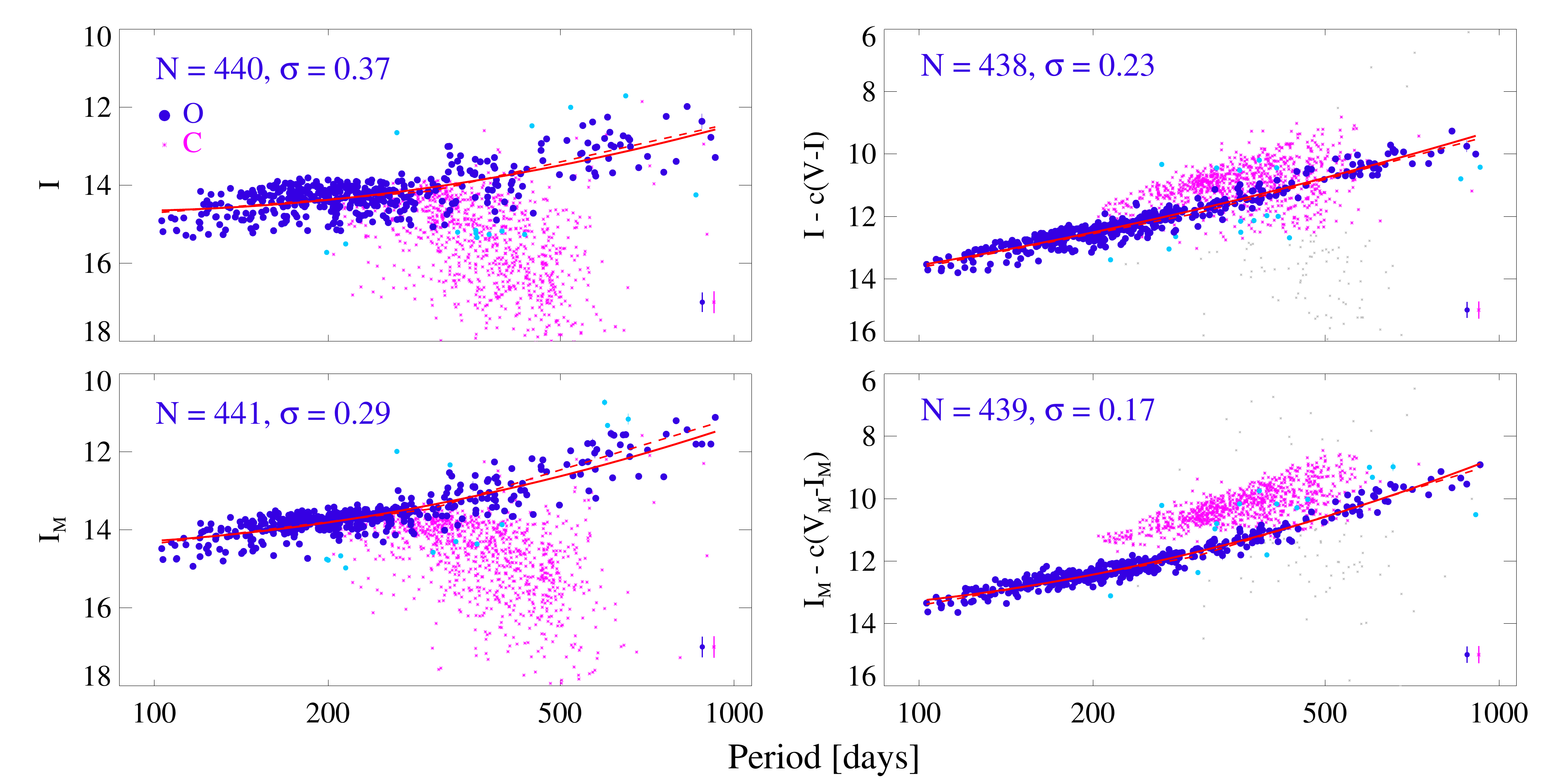}
\caption{PL and PLC relations for Miras at mean- (top) and max-light (bottom) in the LMC using OGLE-III data. The solid/dashed line displays the best-fitting quadratic/non-linear regression model. 
In case of a two-slope linear regression, the break period is adopted at 300 days. Small cyan circles and grey dots are outliers excluded from the O- and C-rich regression analysis, 
respectively. In the case of PLC relations, the color-coefficient ($c$) is different for O- and C-rich Miras (see Table~\ref{tbl:ogl_mc}). Representative $\pm 5\sigma$ errors are shown in each panel. The final number of O-rich Miras used in the quadratic regression and corresponding dispersion are provided at the top of each panel. \label{fig:ogl_lmc_plc}}
\end{figure*}

\begin{figure}
\epsscale{1.18}
\plotone{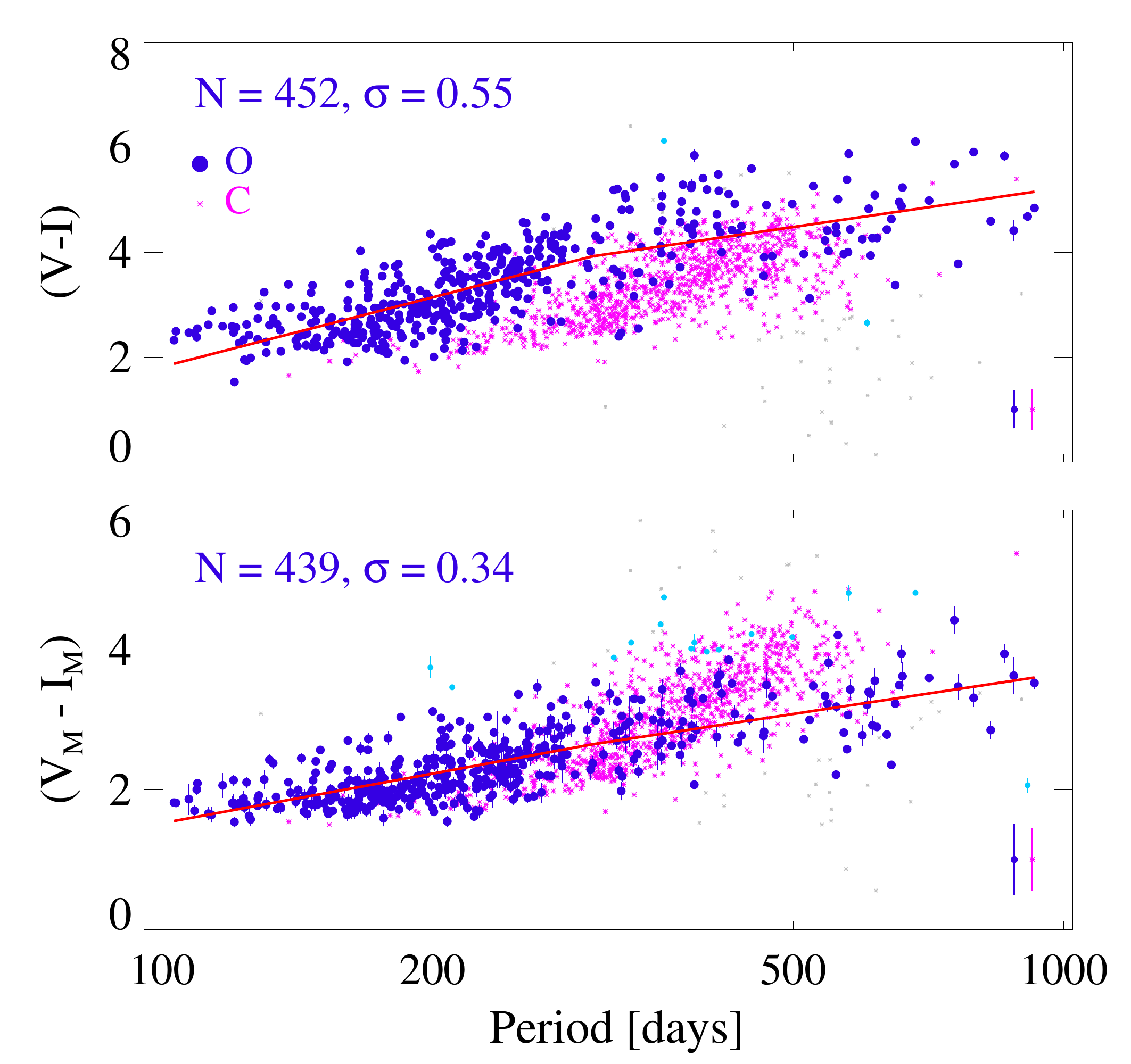}
\caption{PC relations for Miras at mean- (top) and max-light (bottom) in the LMC using OGLE-III data. The solid line represents the best-fitting non-linear regression model with 
a break period at 300 days. Symbols are the same as in Fig.~\ref{fig:ogl_lmc_plc}. \label{fig:pc_lmc_ogl}}
\end{figure}

\begin{figure}
\epsscale{1.18}
\plotone{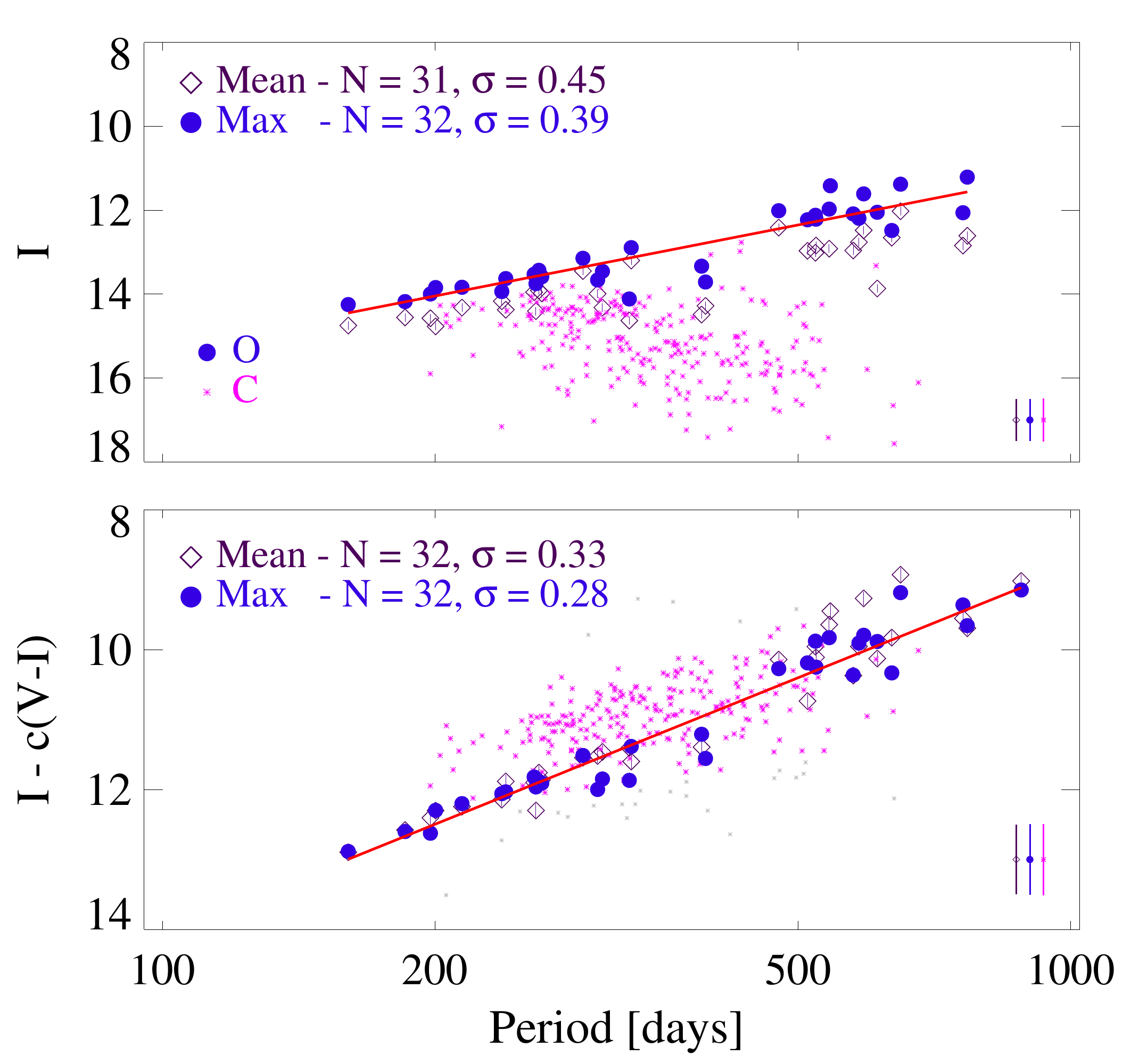}
\caption{PL and PLC relations for Miras in the SMC using OGLE-III data. The diamonds and circles represent the mean- and max-light O-rich Mira PL/PLC relations, respectively. 
Only the max-light PL/PLC relation for C-rich Miras (dots) is shown. The solid line represents a linear regression over the entire period-range. 
The rest is the same as in Fig.~\ref{fig:ogl_lmc_plc}. \label{fig:ogl_smc_plc}}
\end{figure}

\begin{figure*}
\epsscale{1.2}
\plotone{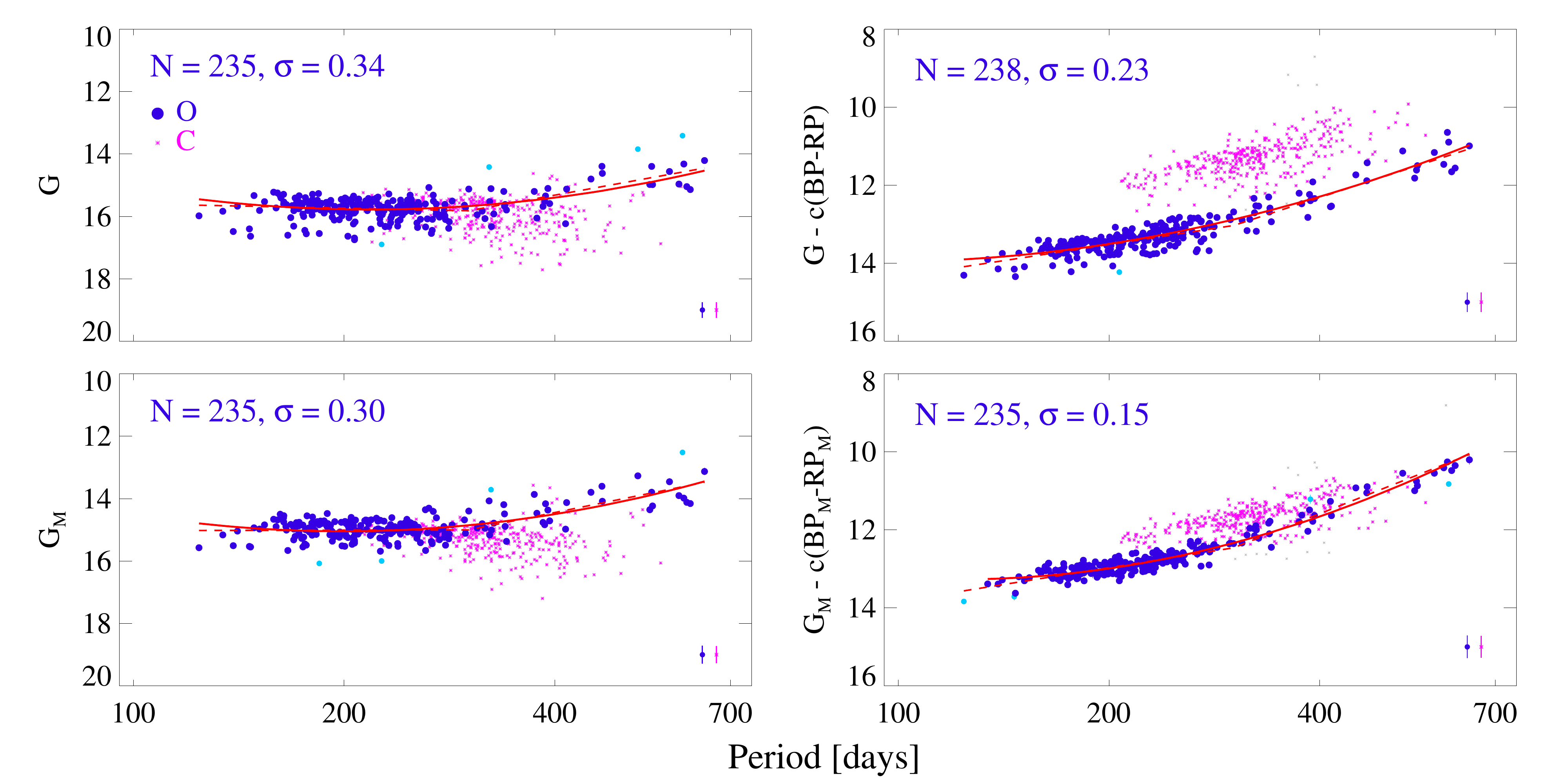}
\caption{As Fig.~\ref{fig:ogl_lmc_plc} but using the {\it Gaia} $G$-filter and $(BP-RP)$ colors. \label{fig:g_lmc_plc}}
\end{figure*}

\subsection{PL and PLC relations for OGLE Miras}
The GPR model-fits are used to estimate $VI$-band magnitudes at the same-epoch for OGLE-III Miras and to derive PL and PLC relations at mean and max-light. Fig.~\ref{fig:ogl_lmc_plc}
shows the $I$-band PL and PLC relations for Miras in the LMC. The max-light PLR for O-rich Miras in the LMC displays smaller scatter than the mean-light relation. 
C-rich Miras do not display any linear relation with their primary pulsation period in optical bands as they exhibit varying levels of circumstellar extinction \citep{ita2011}. 
Adding the color-term results in a significant decrease in the scatter for both O- and C-rich Miras at mean and max-light. The results of the linear (single or two-slope model) and the quadratic 
regressions to the PL and PLC relations are listed in Table~\ref{tbl:ogl_mc}. 

The max-light PL and PLC relations exhibit scatter which is up to $\sim 30\%$ smaller than their mean-light
counterparts. The results for the max-light PLR also hold if we only take the minimum numerical magnitude of the light curve as the max-light magnitude in the $I$-band. 
In this case, the dispersion in O-rich Mira PLR ($N=440,~\sigma=0.27$~mag) is similar to the dispersion when we use our adopted definition (see, Table~\ref{tbl:ogl_mc}).
The residuals of Mira PLRs correlate with colors, as expected, but do not show any dependence on their periods. However, the 
residuals from the mean-light PL and PLC relations are larger than the max-light residuals at both extremes of the $I$-band amplitude distribution.
At minimum-light, the O-rich Miras fall well-above the faint-limit of OGLE-III, and their PLC relations exhibit a dispersion ($N=441,~\sigma=0.47$~mag) almost three times 
the scatter in the max-light relations. 

The period-color relations for Miras at mean- and max-light are shown in Fig.~\ref{fig:pc_lmc_ogl}. For long-period ($P>300$ days) O-rich Miras, the period-color relation
becomes shallower both at mean- and max-light, and the scatter in the relation over the entire period-range is smaller at max-light. If the extinction corrections are applied,
no significant changes are noted in the scatter in these relations. Assuming that the scatter in these relations is intrinsic, the temperature variations at max-light are
significantly smaller than at mean-light for a group of Miras with similar periods: this results in the tighter PLC relation at the brightest epochs. 

Fig.~\ref{fig:ogl_smc_plc} displays the PL and PLC relations at max-light for Miras in the SMC. The number of O-rich Miras is very small in the SMC \citep[see,][]{boyer2011, goldman2018} but 
they also exhibit significantly smaller dispersions in these relations at max-light when compared to their counterparts at mean-light. Note that the geometric extension 
and the line-of-sight depth of the SMC are much larger than for the LMC \citep{subramanian2015, muraveva2018a}, which contributes to the larger scatter in PLRs. Therefore, single-slope models
are fitted to the SMC PL and PLC relations, considering the dispersion and smaller number statistics. The color dependence of these PLRs for C-rich Miras 
is similar in both Magellanic Clouds. On the other hand, O-rich Miras in the SMC display a greater color dependence in the PLC relations than Miras 
in the LMC, albeit also with greater uncertainty. Note that O-rich stars produce dust less efficiently in low-metallicity environments while 
C-rich Miras are as efficient as their higher-metallicity LMC counterparts \citep{sloan2010,mcdonald2010,sloan2012,sloan2016}. This difference in the dust properties for O-rich Miras 
due to metallicity differences can contribute to the changes in the color coefficient of the PLC relation but the effects of molecular absorption and the temperature 
variations in O-rich Miras may also depend on metallicity.

\begin{figure*}
\epsscale{1.2}
\plotone{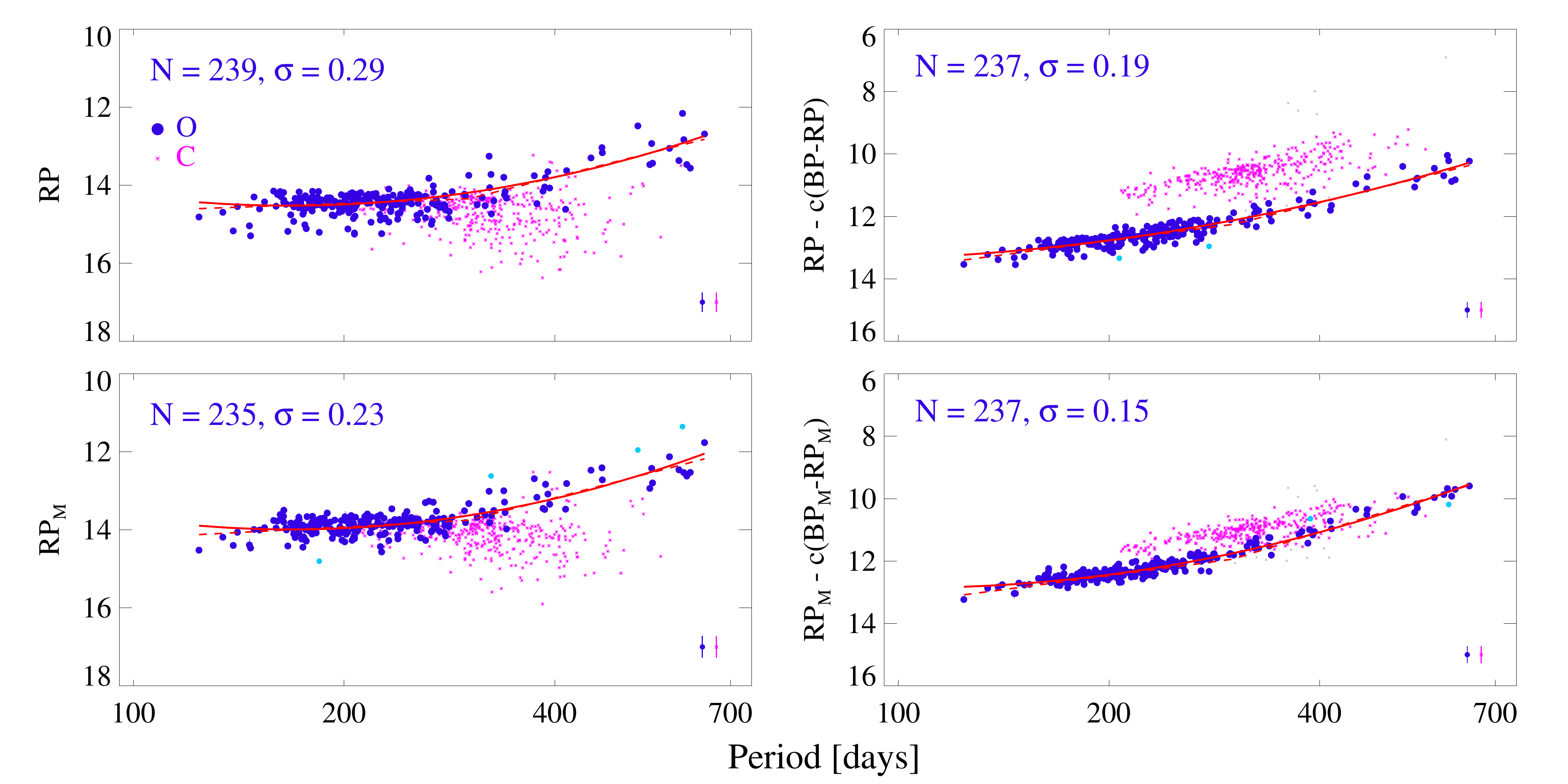}
\caption{As Fig.~\ref{fig:ogl_lmc_plc} but using the {\it Gaia} $RP$-filter and $(BP-RP)$ colors. \label{fig:rp_lmc_plc}}
\end{figure*}

To validate the statistical significance of the decrease in the dispersion of the PL and PLC relations at max-light, a parametric F-test 
is adopted to compare the variances in the mean- and max-light relations. 
The smaller scatter in the max-light relations may also be due to the smaller number of stars in the final fit, for example, due to iterative outlier clipping algorithm. Under the null 
hypothesis that the variances are similar in the mean- and max-light relations, the F-test results suggest that the reduction in the dispersion is statistically significant at 
the $95\%$ confidence interval. The probability of acceptance of the null hypothesis is smaller than $0.05$~dex in all cases.

\subsection{PL and PLC relations for Gaia Miras} \label{sec:plc_gaia}

\begin{figure}
\epsscale{1.2}
\plotone{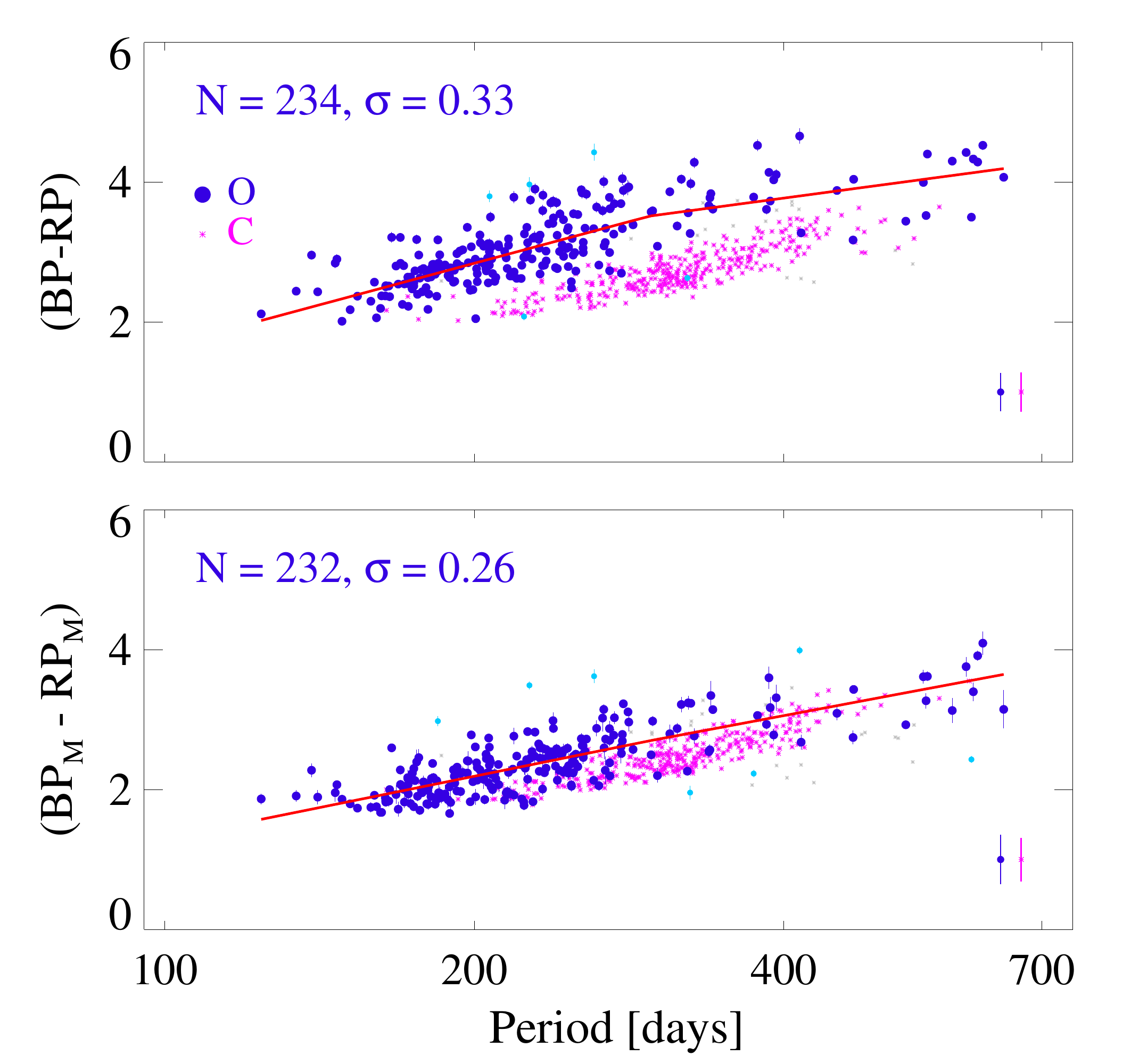}
\caption{PC relations for Miras at mean- (top) and max-light (bottom) in the LMC using {\it Gaia} data. The solid line represents a best-fitting non-linear regression model with 
a break period at 300 days. Symbols are the same as in Fig.~\ref{fig:ogl_lmc_plc}. \label{fig:pc_lmc_gaia}}
\end{figure}

\begin{deluxetable*}{llrrrrrrrr}
\tablecaption{The PL and PLC relations for Miras in the Magellanic Clouds using the {\it Gaia} data. \label{tbl:gaia_mc}}
\tabletypesize{\footnotesize}
\tablewidth{0pt}
\tablehead{\colhead{MC} & \colhead{Band} & \colhead{O/C}& \colhead{$a$} & \colhead{$b$} & \colhead{$b_l$} & \colhead{$c$} & \colhead{$\sigma$} & \colhead{$N_i$}  & \colhead{$N_f$}}
\startdata
\multicolumn{10}{c}{{Quadratic regression}}\\
\cline{1-10}
LMC               & $G$& O &      15.67$\pm$0.03      &	$-$1.54$\pm$0.17      &	$-$5.50$\pm$0.78      &                 ---      &       0.34&          239&          235\\
                & $G_M$& O &      14.85$\pm$0.03      &	$-$2.16$\pm$0.17      &	$-$6.02$\pm$0.79      &                 ---      &       0.30&          239&          235\\
          & $G,(BP-RP)$& O &      12.88$\pm$0.15      &	$-$4.27$\pm$0.17      &	$-$4.18$\pm$0.55      &       0.79$\pm$0.04      &       0.23&          239&          238\\
    & $G_M,(BP_M-RP_M)$& O &      12.33$\pm$0.09      &	$-$4.77$\pm$0.12      &	$-$6.09$\pm$0.43      &       0.93$\pm$0.03      &       0.15&          239&          235\\
                 & $RP$& O &      14.19$\pm$0.03      &	$-$2.61$\pm$0.13      &	$-$5.12$\pm$0.63      &                 ---      &       0.29&          239&          239\\
               & $RP_M$& O &      13.63$\pm$0.02      &	$-$2.84$\pm$0.13      &	$-$5.61$\pm$0.57      &                 ---      &       0.23&          239&          235\\
         & $RP,(BP-RP)$& O &      12.12$\pm$0.13      &	$-$4.27$\pm$0.14      &	$-$3.62$\pm$0.46      &       0.60$\pm$0.04      &       0.19&          239&          237\\
   & $RP_M,(BP_M-RP_M)$& O &      11.75$\pm$0.09      &	$-$4.82$\pm$0.12      &	$-$5.23$\pm$0.39      &       0.69$\pm$0.03      &       0.15&          239&          237\\
\cline{1-10}
\multicolumn{10}{c}{Linear regression}\\
\cline{1-10}
LMC               & $G$& O &      15.82$\pm$0.05      &       0.46$\pm$0.27      &	$-$4.08$\pm$0.37      &                 ---      &       0.34&          239&          236\\
                & $G_M$& O &      15.00$\pm$0.04      &	$-$0.05$\pm$0.25      &	$-$4.55$\pm$0.39      &                 ---      &       0.30&          239&          235\\
          & $G,(BP-RP)$& O &      13.03$\pm$0.16      &	$-$2.75$\pm$0.25      &	$-$5.85$\pm$0.26      &       0.78$\pm$0.04      &       0.23&          239&          238\\
    & $G_M,(BP_M-RP_M)$& O &      12.46$\pm$0.10      &	$-$2.87$\pm$0.17      &	$-$7.13$\pm$0.22      &       0.93$\pm$0.03      &       0.16&          239&          238\\
                 & $RP$& O &      14.35$\pm$0.04      &	$-$0.65$\pm$0.21      &	$-$4.55$\pm$0.28      &                 ---      &       0.28&          239&          237\\
               & $RP_M$& O &      13.75$\pm$0.03      &	$-$0.96$\pm$0.19      &	$-$4.69$\pm$0.28      &                 ---      &       0.23&          239&          236\\
         & $RP,(BP-RP)$& O &      12.25$\pm$0.13      &	$-$2.98$\pm$0.22      &	$-$5.62$\pm$0.21      &       0.59$\pm$0.04      &       0.19&          239&          237\\
   & $RP_M,(BP_M-RP_M)$& O &      11.93$\pm$0.09      &	$-$2.99$\pm$0.16      &	$-$7.05$\pm$0.21      &       0.66$\pm$0.03      &       0.15&          239&          238\\
          & $G,(BP-RP)$& C &      11.40$\pm$0.24      &	$-$3.73$\pm$0.38      &                 ---      &       1.69$\pm$0.09      &       0.35&          359&          351\\
    & $G_M,(BP_M-RP_M)$& C &      11.75$\pm$0.20      &	$-$3.10$\pm$0.32      &                 ---      &       1.44$\pm$0.08      &       0.28&          359&          346\\
         & $RP,(BP-RP)$& C &      10.69$\pm$0.23      &	$-$3.86$\pm$0.36      &                 ---      &       1.50$\pm$0.09      &       0.33&          359&          351\\
   & $RP_M,(BP_M-RP_M)$& C &      11.09$\pm$0.18      &	$-$3.40$\pm$0.29      &                 ---      &       1.25$\pm$0.07      &       0.26&          359&          345\\
\cline{1-10}
SMC               & $G$& O &      15.18$\pm$0.08      &	$-$4.02$\pm$0.39      &                 ---      &                 ---      &       0.38&           25&           25\\
                & $G_M$& O &      14.36$\pm$0.05      &	$-$4.55$\pm$0.28      &                 ---      &                 ---      &       0.27&           25&           25\\
          & $G,(BP-RP)$& O &      13.06$\pm$0.40      &	$-$5.17$\pm$0.34      &                 ---      &       0.73$\pm$0.14      &       0.26&           25&           25\\
  & $G_M,(BP_M- RP_M)$& O &      12.71$\pm$0.39      &	$-$5.35$\pm$0.29      &                 ---      &       0.78$\pm$0.18      &       0.20&           25&           25\\
                 & $RP$& O &      13.93$\pm$0.06      &	$-$4.34$\pm$0.32      &                 ---      &                 ---      &       0.31&           25&           25\\
               & $RP_M$& O &      13.31$\pm$0.05      &	$-$4.86$\pm$0.25      &                 ---      &                 ---      &       0.24&           25&           25\\
         & $RP,(BP-RP)$& O &      12.44$\pm$0.37      &	$-$5.15$\pm$0.31      &                 ---      &       0.51$\pm$0.13      &       0.24&           25&           25\\
 & $RP_M,(BP_M- RP_M)$& O &      12.17$\pm$0.38      &	$-$5.42$\pm$0.28      &                 ---      &       0.54$\pm$0.18      &       0.20&           25&           25\\
          & $G,(BP-RP)$& C &      11.56$\pm$0.46      &	$-$3.79$\pm$0.50      &                 ---      &       1.72$\pm$0.16      &       0.37&           77&           76\\
  & $G_M,(BP_M- RP_M)$& C &      10.86$\pm$0.30      &	$-$3.73$\pm$0.34      &                 ---      &       1.90$\pm$0.12      &       0.24&           77&           73\\
         & $RP,(BP-RP)$& C &      10.87$\pm$0.42      &	$-$3.62$\pm$0.47      &                 ---      &       1.53$\pm$0.15      &       0.34&           77&           76\\
 & $RP_M,(BP_M- RP_M)$& C &      10.33$\pm$0.30      &	$-$3.74$\pm$0.33      &                 ---      &       1.66$\pm$0.12      &       0.23&           77&           73\\
\cline{1-10}
\enddata
\tablecomments{The coefficients ($a, b, b_l, c$) are defined in Section~\ref{sec:data}. $\sigma$: dispersion (mag). $N_i$: initial number of sources. $N_f$: final number after excluding outliers.}
\end{deluxetable*}

The {\it Gaia} mission has provided one of the largest samples of LPVs \citep{mowlavi2018}, which has enabled us to study 
PL and PLC relations with simultaneous multi-band precise space-based photometry. This LPV sample is restricted to Miras with periods 
less than 670 days, owing to the observation duration of the {\it Gaia} data. Further, some of the epochs in the light curves that were 
flagged by the photometry and variability processing by the {\it Gaia} team are also excluded 
from this analysis. Although the magnitudes corresponding to most of those epochs fall on the light curves \citep{mowlavi2018}, these were
not included in our adopted criterion of the minimum number of 15 epochs per light curve. The mean- and max-light magnitudes in 
$G$, $BP$, and $RP$ filters are obtained for epochs of simultaneous photometry in all three filters.

Figs.~\ref{fig:g_lmc_plc} and \ref{fig:rp_lmc_plc} show the PL and PLC relations for Miras in the LMC based on $G$ and $RP$ band magnitudes and $(BP-RP)$ colors, respectively. The
results of the regression analysis are tabulated in Table~\ref{tbl:gaia_mc}. Similar to the traditional $VI$ filters, PL and PLC relations at max-light display significantly 
smaller dispersions than at mean-light in the {\it Gaia} filters as well. The scatter in the max-light PLC relations for O-rich Miras is comparable to that for the PLRs in the NIR $JH$-bands in 
the LMC \citep{yuan2017b}. Fig.~\ref{fig:pc_lmc_gaia} presents period-color relations for Miras using {\it Gaia} $(BP-RP)$ colors.
The O- and C-rich Miras display two distinct relations at mean-light while the temperature variations are smaller at max-light. 
The max-light colors for C-rich Miras are very similar to O-rich Miras. The flatter color variation for longer-period O-rich Miras is also seen at mean-light 
similar to that in $(V-I)$ color. 

\begin{figure}
\epsscale{1.2}
\plotone{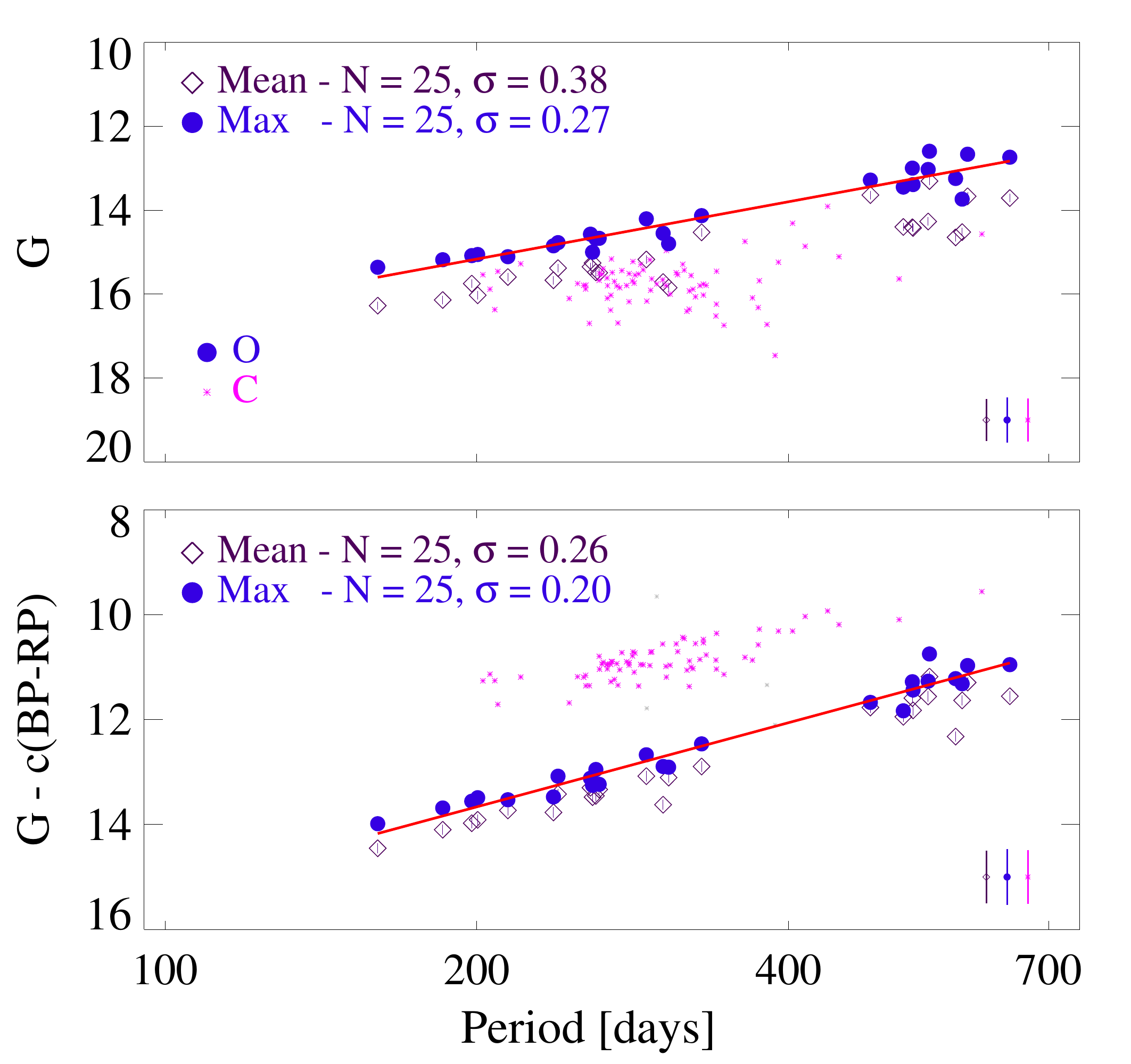}
\caption{PL and PLC relations for Miras in the SMC using {\it Gaia} data. The diamonds and circles represent mean- and max-light O-rich Mira PL/PLC relations, respectively. 
Only max-light PL/PLC relation for C-rich Miras (dots) are shown. The solid line represents a linear regression over the entire period-range. 
The rest is the same as in Fig.~\ref{fig:ogl_lmc_plc}. \label{fig:g_smc_plc}}
\end{figure}

Fig.~\ref{fig:g_smc_plc} displays the PL and PLC relations at max-light for Miras in the SMC. The decrease in the scatter in these relations is consistent 
with the results for Miras in the LMC. The color term for C-rich Miras is almost twice than for O-rich Miras in all optical colors. Assuming similar interstellar extinction, the effect of circumstellar extinction on C-rich Miras may be twice as large as that for O-rich Miras.

\subsection{Quadratic or non-linear relations}

Optical and NIR PL and PLC relations for long-period ($P\gtrsim 300$ days) O-rich Miras exhibit a steeper slope than the shorter period Miras \citep{feast1989, ita2011}. 
The excess luminosity in long-period O-rich Miras possibly occurs due to hot bottom burning \citep[HBB,][]{sackman1992, whitelock2003} episodes in massive AGB 
stars \citep[$\sim 3-5M_\odot$, for details, see][]{karakas2014, marigo2017}. For AGB stars with initial masses between $\sim 3$ and $\sim 5M_\odot$, nuclear
burning at the base of the convective envelope leads to luminosities that are higher than the core-mass-luminosity predictions. The HBB process also affects the surface chemistry 
of AGB stars, and the lower mass limit to initiate these episodes depends on metallicity \citep{marigo2013}. The mass-loss in AGB stars also changes as it evolves towards brighter 
luminosity and longer periods \citep{vassili1993, goldman2017, hofner2018}. Empirically, the exact period at which these effects may cause a change in the slope of the PLRs 
is not well-quantified. \citet{ita2011} found a kink in the PLRs around 400 days. A quadratic relation 
also seems to provide a robust fit to PLRs over the entire period range of O-rich Miras \citep{yuan2017b}.

\begin{deluxetable}{lrrrrr}
\tablecaption{Results of the statistical tests to find a kink in the O-rich Mira PL and PLC relations in the LMC. \label{tbl:plc_test}}
\tabletypesize{\footnotesize}
\tablewidth{0pt}
 \tablehead{\colhead{PL/PLC} & \colhead{SIC} & \colhead{RW}& \colhead{TM}& \colhead{FT} & \colhead{$p(f)$}}
\startdata
\cline{1-6}
---	&	\multicolumn{4}{c}{$\log P_b$}	&	---\\
\cline{1-6}
                     $I,(V-I)$&       2.37&       2.42&       2.35&       2.35&       0.51\\
               $I_M,(V_M-I_M)$&       2.45&       2.49&       2.43&       2.48&       0.78\\
                   $G,(BP-RP)$&       2.44&       2.43&       2.48&       2.45&       0.36\\
             $G_M,(BP_M-RP_M)$&       2.40&       2.39&       2.48&       2.46&       0.88\\
                  $RP,(BP-RP)$&       2.43&       2.41&       2.47&       2.43&       0.32\\
            $RP_M,(BP_M-RP_M)$&       2.40&       2.42&       2.48&       2.42&       0.84\\
                           $J$&       2.50&       2.50&       2.53&       2.51&       0.01\\
                         $J_M$&       2.55&       2.47&       2.48&       2.54&       0.23\\
                           $K$&       2.52&       2.51&       2.49&       2.53&       0.19\\
                         $K_M$&       2.54&       2.48&       2.50&       2.54&       0.01\\
\cline{1-6}	
\enddata
\tablecomments{SIC: Schwarz Information Criterion used to estimate the break period in a two-slope model.\\
RW: Random-walk, TM: The Testimator,  FT: F-test\\
$p(f)$: Probability of acceptance of the null hypothesis that the variance of a quadratic and a linear relation for $P<400$ days are equal.}
\end{deluxetable}

The statistical significance of linear, non-linear or quadratic models is investigated using several methods: Schwarz Information Criterion (SIC), Random Walk (RW), 
Testimator (TM) and F-test \citep[FT, see ][for mathematical details]{kanbur2007, bhardwaj2016b}. In the case of a non-linear regression model, a break period is varied between
200 and 500 days in steps of $\log P=0.01$~day. The SIC is used to compare the quadratic and two-slope regression model by maximizing the likelihood function and the model 
with the smallest SIC value is adopted as the preferred model. A break-period is obtained from the two-slope regression model with the smallest SIC value. The RW method gives the probability of 
the acceptance of the null hypothesis that a non-linear regression is the best-model, and the break-period corresponds to the maximum-probability.
The TM compares the slopes for different subsets of the data under consideration and provides a period-range where the PLRs depart from linearity.
The F-test compares the variances of the quadratic and non-linear regression models under the null hypothesis that the two variances are equal. 
The period corresponding to the model with the largest F-value is adopted as the break-period. Finally, a quadratic relation is compared with a 
linear relation for O-rich Miras with periods smaller than $400$ days. The probability $p(f) < 0.05$ suggests that the two variances are not equal.

Table~\ref{tbl:plc_test} summarizes the results of the statistical tests to determine a break in the PL and PLC relations for O-rich Miras in the LMC. 
Since these methods are sensitive to the dispersion of the underlying relation, optical PLC relations and NIR PLRs are used to determine the break period. 
Non-linear regression is the best-model according to SIC. The break-period varies between $2.35 <\log P < 2.55$ days at different wavelengths for all
statistical methods. The F-test suggests that the variances of the quadratic and two-slope models are statistically similar. Further, the variances 
of a quadratic model and a linear relation for $P < 400$ days are also similar in most cases. Therefore, we adopt a break period of 300 days 
($\sim \log P = 2.48$) as the optimal period at which the change in the slope of PL and PLC relations occurs for O-rich Miras. The break-period possibly 
shifts to longer periods ($\sim$350 days, $\sim \log P = 2.54$ days) at NIR wavelengths. NIR PLRs are discussed in detail in Section~\ref{sec:mira_nir}.

\section{Stability of maximum-light for Miras}
\label{sec:mira_max}

\subsection{Photometric variability at max-light}

Optical photometry provides clear evidence that Miras display significantly smaller scatter in the PLRs at max-light than at mean-light. The pulsation properties and brightness variations 
of Miras are not strictly periodic and vary from cycle-to-cycle. Therefore, it is important to investigate the stability of max-light over different pulsation
cycles. However, apart from OGLE $I$-band time-series observations, there is a lack of continuous photometry for Miras in the LMC, and even OGLE data do not necessarily cover the epochs of max-light
during all pulsation cycles. High-precision photometry and sub-hour cadence from {\it Kepler} can be extremely useful to test the stability of max-light over different pulsation cycles. 
\citet{banyai2013} analyzed variability in $M$-giants from {\it Kepler} and identified a dozen long-period Mira-like variables. 

We estimate the scatter in the max-light magnitudes and median magnitudes for Miras over different pulsation cycles. For OGLE Miras, the sample is restricted to those stars for which GPR predicts 
max-light magnitudes with a precision better than three times their median uncertainty, and the time coverage is equivalent to the period of each Mira. The top panel of Fig.~\ref{fig:sigma_mag} displays the comparison of the scatter in the median and max-light magnitudes for OGLE and {\it Kepler} Mira candidates. 
When the scatter in both sets of magnitudes is smaller than 0.4~mag, max-light magnitudes seem to be more stable for most O-rich Miras. 
C-rich Miras often exhibit long-term trends and thus display large scatter both at median- and max-light. The bottom panel of Fig.~\ref{fig:sigma_mag} presents 
the variation in min-light magnitudes versus the scatter in median magnitudes. For a majority of the stars, min-light magnitudes display significantly larger scatter than median 
magnitudes over multiple pulsation cycles. The scatter in max-light magnitudes is systematically smaller than in min-light magnitudes with an average offset of 0.06 mag and the
largest offset of 0.3~mag. 
High-precision photometry of {\it Kepler} Miras also displays smaller magnitude variations at max-light than at min-light.
Similar magnitude variations are seen at max- and mean-light when the dispersion over multiple pulsation cycles in both sets of the magnitudes is large ($>0.4$~mag). 
The lack of accurate parallaxes 
for the {\it Kepler} Miras and solar-neighborhood Miras with high-cadence photometry precludes us from quantifying the impact of scatter in mean- and max-light magnitudes on the PLRs. 
Note that accurate {\it Gaia} parallaxes for these Miras are not available presumably due to large astrometric errors that are caused by the motion of the stellar photocentre \citep{chiavassa2018}.
Miras also exhibit up to 40\% variation in their stellar diameters \citep[$H$-band,][]{lacour2009}, and angular diameters can be larger than the parallax measurements. 
Further, Miras display large magnitude and color variations along the pulsation cycle and chromaticity corrections are required at each epoch to estimate accurate parallaxes \citep{mowlavi2018}.
In addition to precise distances, extinction and metallicity measurements are also needed to calibrate PLRs for Miras and to investigate the impact of cycle-to-cycle variations in max-light
magnitudes on their PLRs.

\begin{figure}
\centering
  \begin{tabular}{@{}c@{}}
    \includegraphics[width=.48\textwidth,trim={0.2cm 0.2cm 0cm 0.2cm},clip]{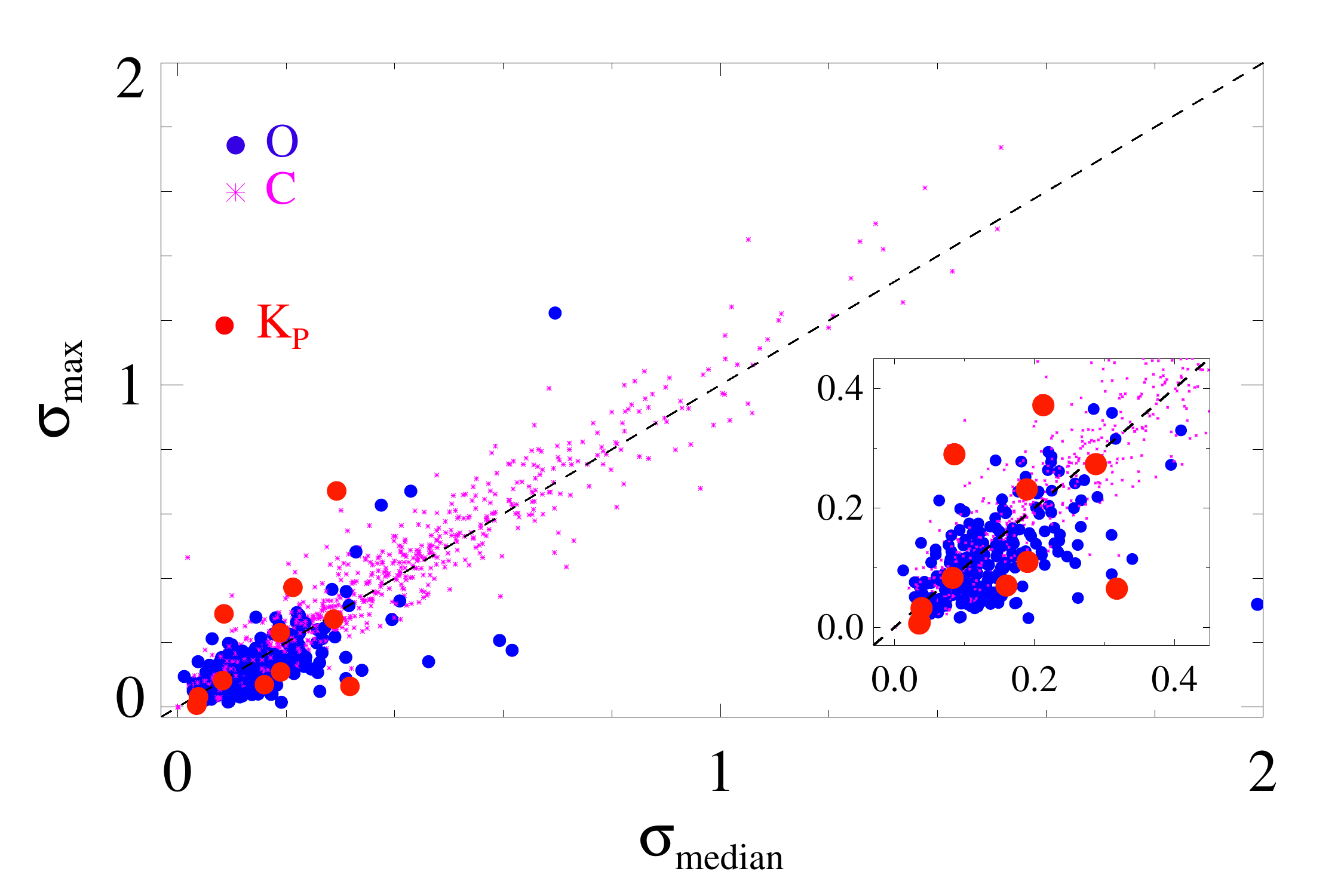} \\
    \includegraphics[width=.48\textwidth,trim={0.2cm 0.2cm 0cm 0.8cm},clip]{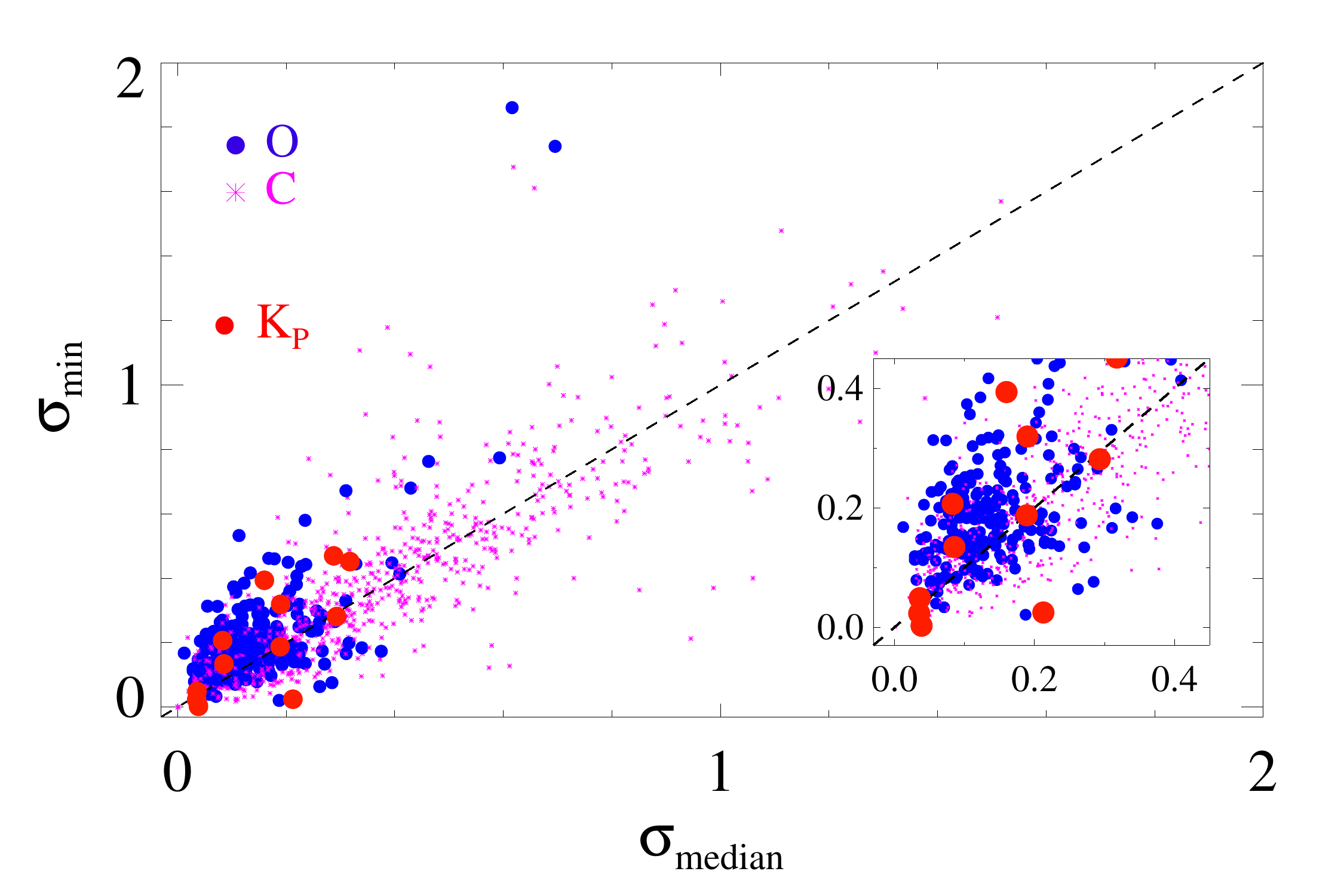} \\
  \end{tabular}
 \caption{{\it Top panel:} Comparison of the scatter in median and max-light magnitudes obtained over multiple pulsation cycles for {\it Kepler} Mira candidates 
(red) and for OGLE sources. {\it Bottom panel:} As above but versus min-light magnitudes. \label{fig:sigma_mag}}
\label{fig:sigma_mag}
\end{figure}

\begin{figure*}
\centering
  \begin{tabular}{@{}cc@{}}
    \includegraphics[width=.5\textwidth]{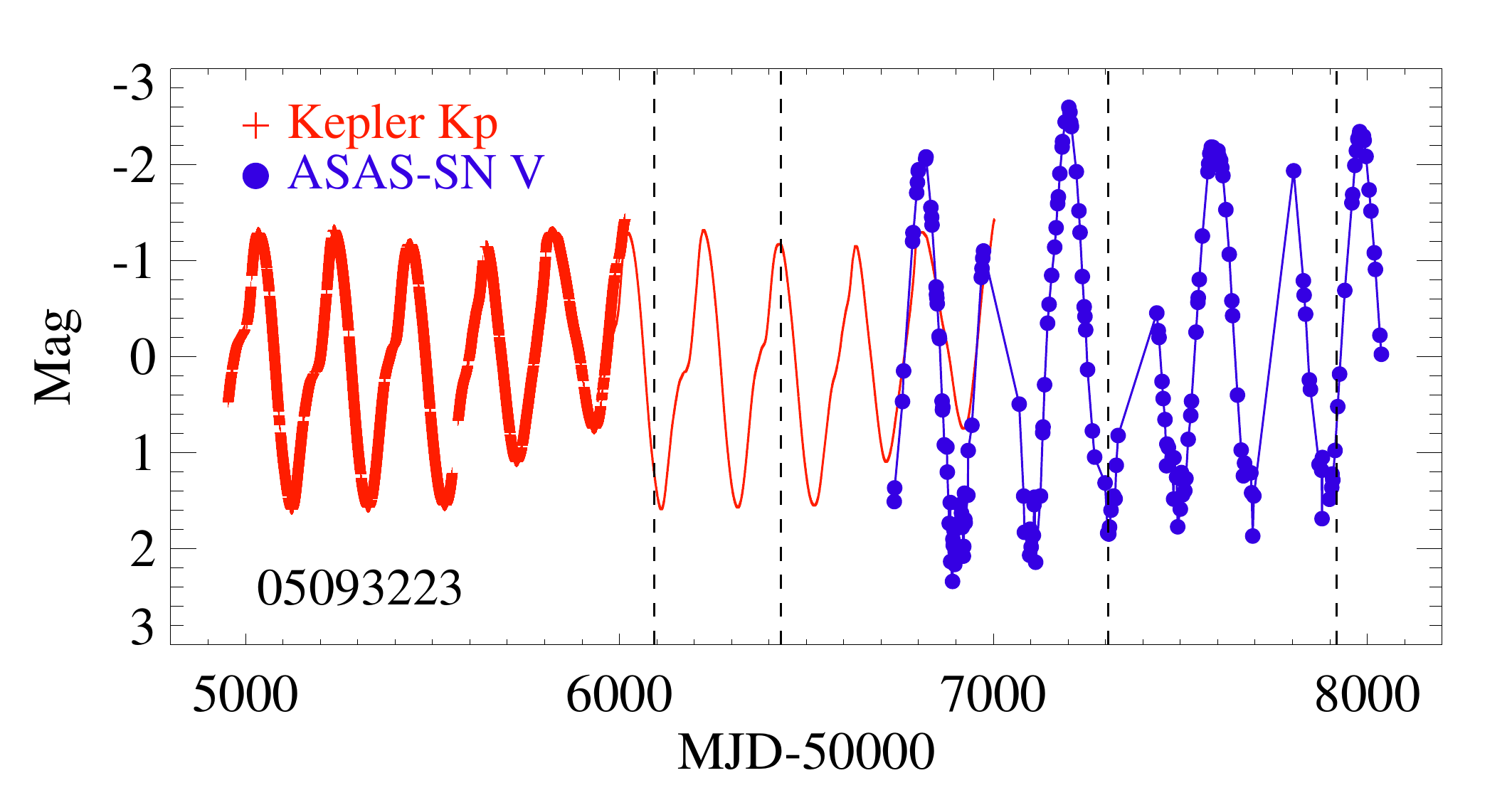} &
    \includegraphics[width=.5\textwidth]{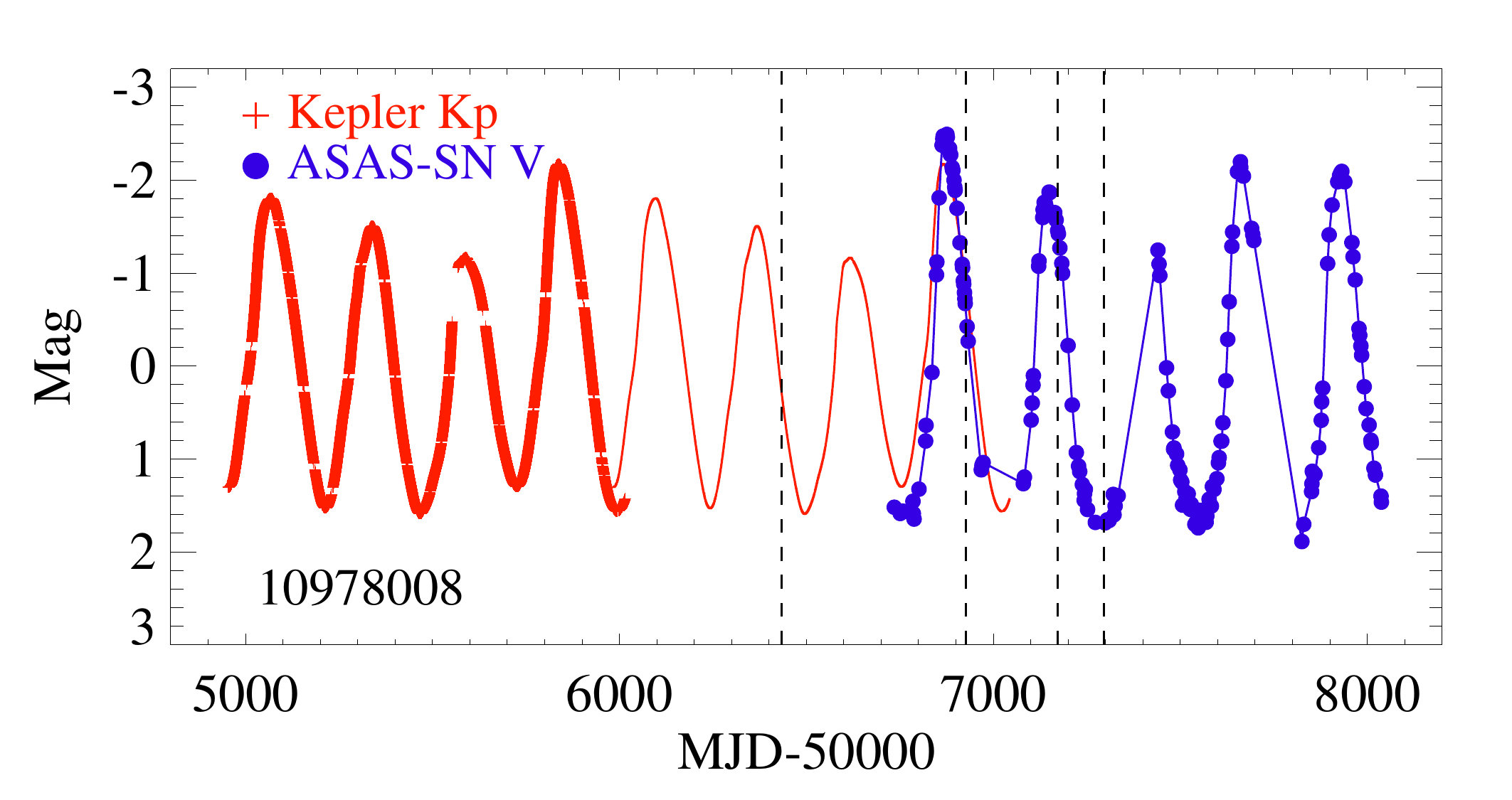} \\
    \includegraphics[width=.5\textwidth,trim={0cm 0.0cm 0cm 0.5cm},clip]{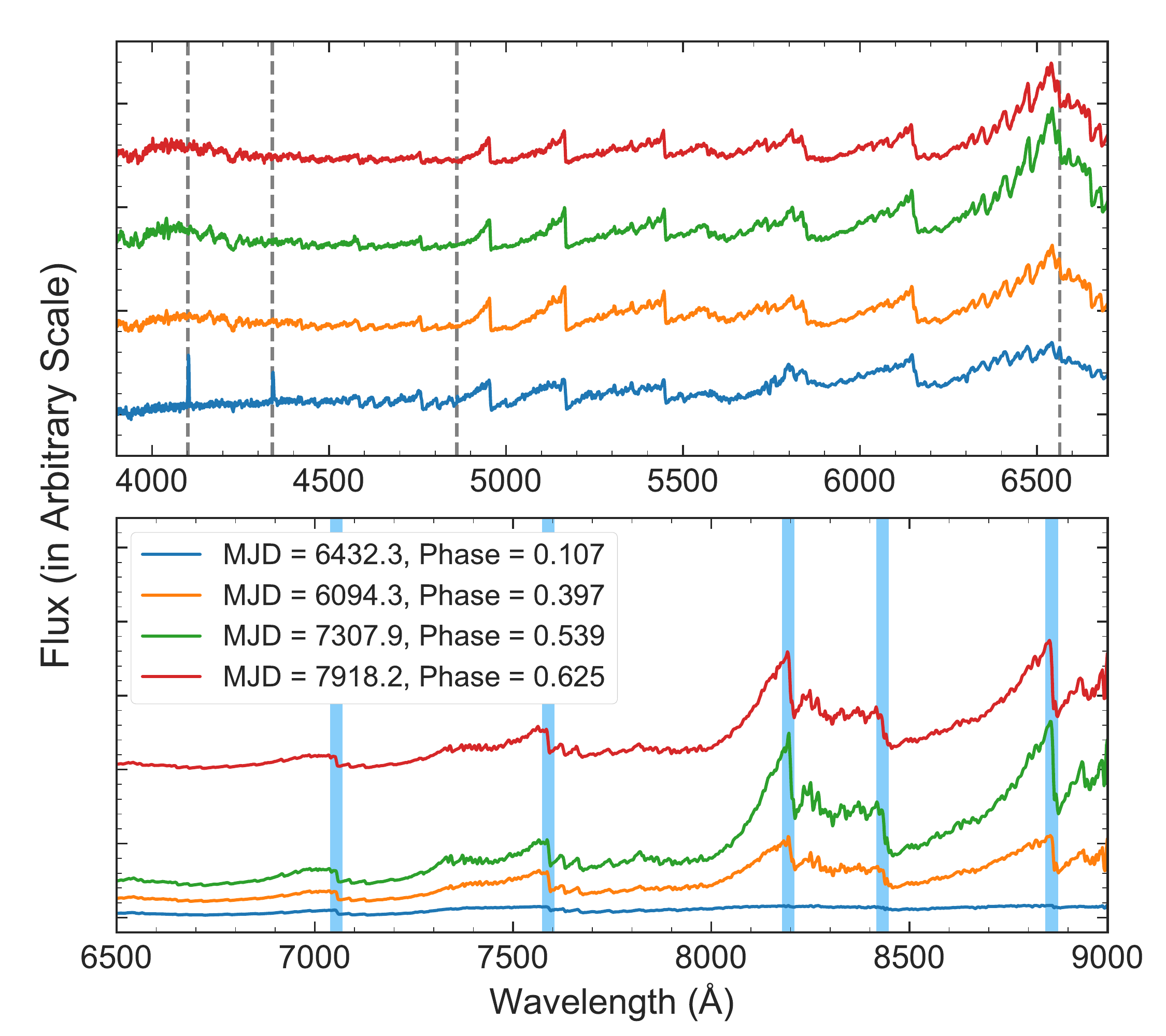} &
    \includegraphics[width=.5\textwidth,trim={0cm 0.0cm 0cm 0.5cm},clip]{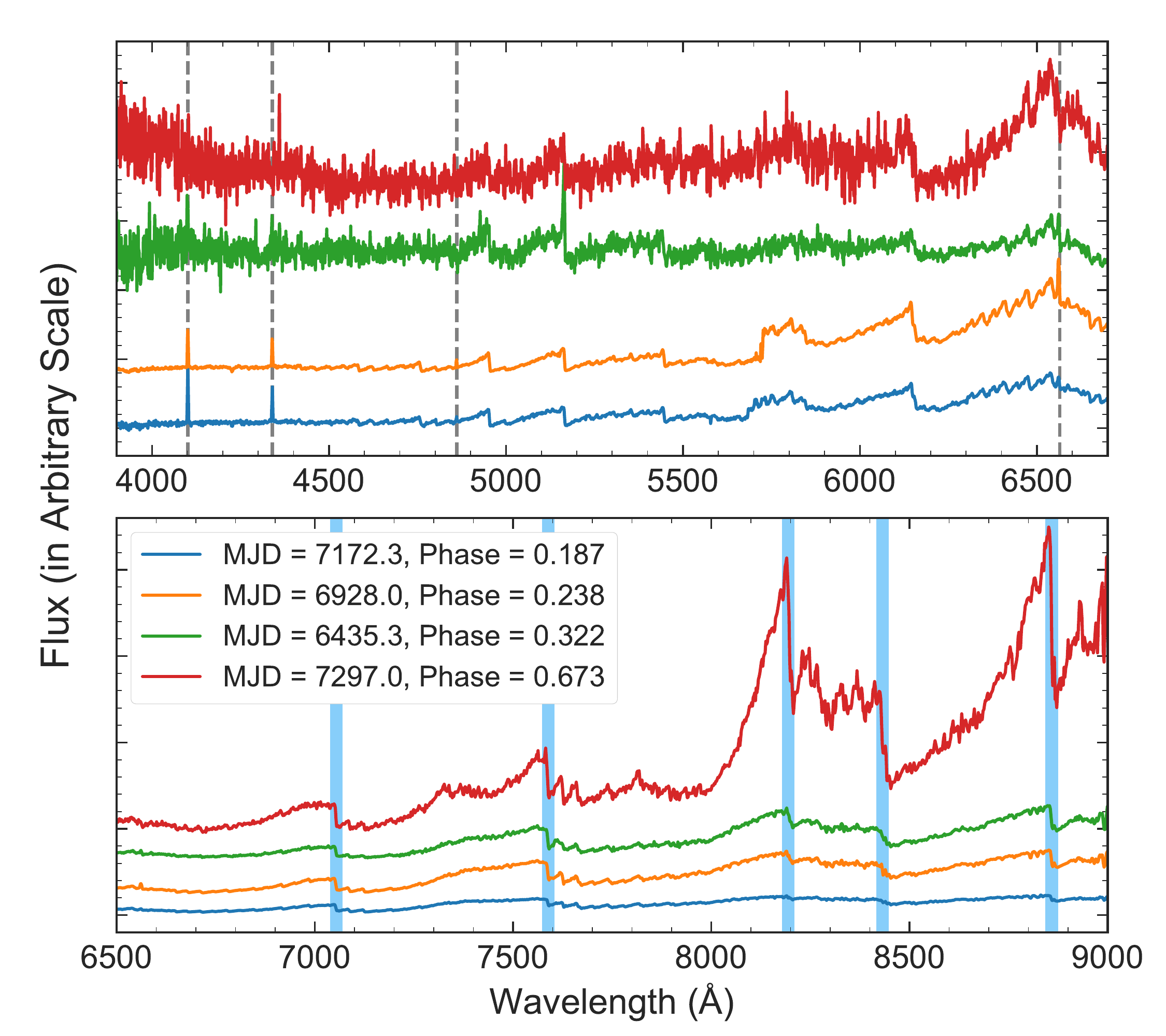} \\
  \end{tabular}
 \caption{{\it Top panels:} Light curves of two {\it Kepler} Mira candidates in the $K_P$ (red plus symbols) and $V$ (blue circles) bands. The solid line represents 
the offset {\it Kepler} light curve to fill in the missing epochs, which does not fit $V$-band epochs perfectly because of unaccounted phase lag and differences in the 
amplitudes. The light curves are normalized to zero-mean. Vertical dashed lines represent the epochs when the spectra were taken. {\it Middle and bottom panels:} Mira spectra at multiple epochs.
Phase zero corresponds to the epoch of max-light in the $K_P$ band. Each spectrum is normalized to unity at 5550\AA ~for a relative comparison of TiO band strengths and an 
offset is applied to multi epoch spectra for clarity. The Balmer series lines ($\lambda<$7000\AA) are marked with dashed (grey) lines and TiO bands ($\lambda>$7000\AA) 
are represented by shaded (blue) lines.}
 \label{fig:kp_spec}
\end{figure*}

\subsection{Spectral features at max-light}

Mira spectra for {\it Kepler} stars are taken from the Large Sky Area Multi-Object Fiber Spectroscopic Telescope (LAMOST) survey \citep{zhao2012,cui2012}. 
The {\it LAMOST-Kepler} project covered the original
target fields at least twice. Therefore, all Mira candidates have one or more spectra, and a study as a function of the pulsation phase is possible wherever more than one spectrum is available. 
Fig.~\ref{fig:kp_spec} shows the light curves of two {\it Kepler} Mira candidates in the top panels and their LAMOST spectra in the bottom panels. The {\it Kepler} $K_P$ and $V$-band light curves 
are taken from \citet{banyai2013} and All-Sky Automated Survey for Supernovae (ASAS-SN) catalog of variable stars \citep{shappee2014, jayasinghe2018}. At least one spectrum close to min, mean, 
and max-light is available for these two Miras. All spectra display TiO molecular features but no sequential vibrational bands of carbon-bearing molecules are seen. 
Therefore, both Miras are classified as O-rich Miras. The max-light spectra (close to the zero phase which corresponds to the max of the $K_P$-band) for {\it Kepler} Miras 
exhibit strong Balmer line emission. Further, both Miras display a Balmer increment at max-light and the strength of TiO bands increases from max- to min-light. 
\citet{yao2017} showed that as an AGB star cools down from its warmest phase, the strength of TiO bands increases gradually and thus more flux is absorbed 
by lower order Balmer lines \citep{merrill1940}.

\begin{figure*}
\centering
  \begin{tabular}{@{}cc@{}}
    \includegraphics[width=.45\textwidth]{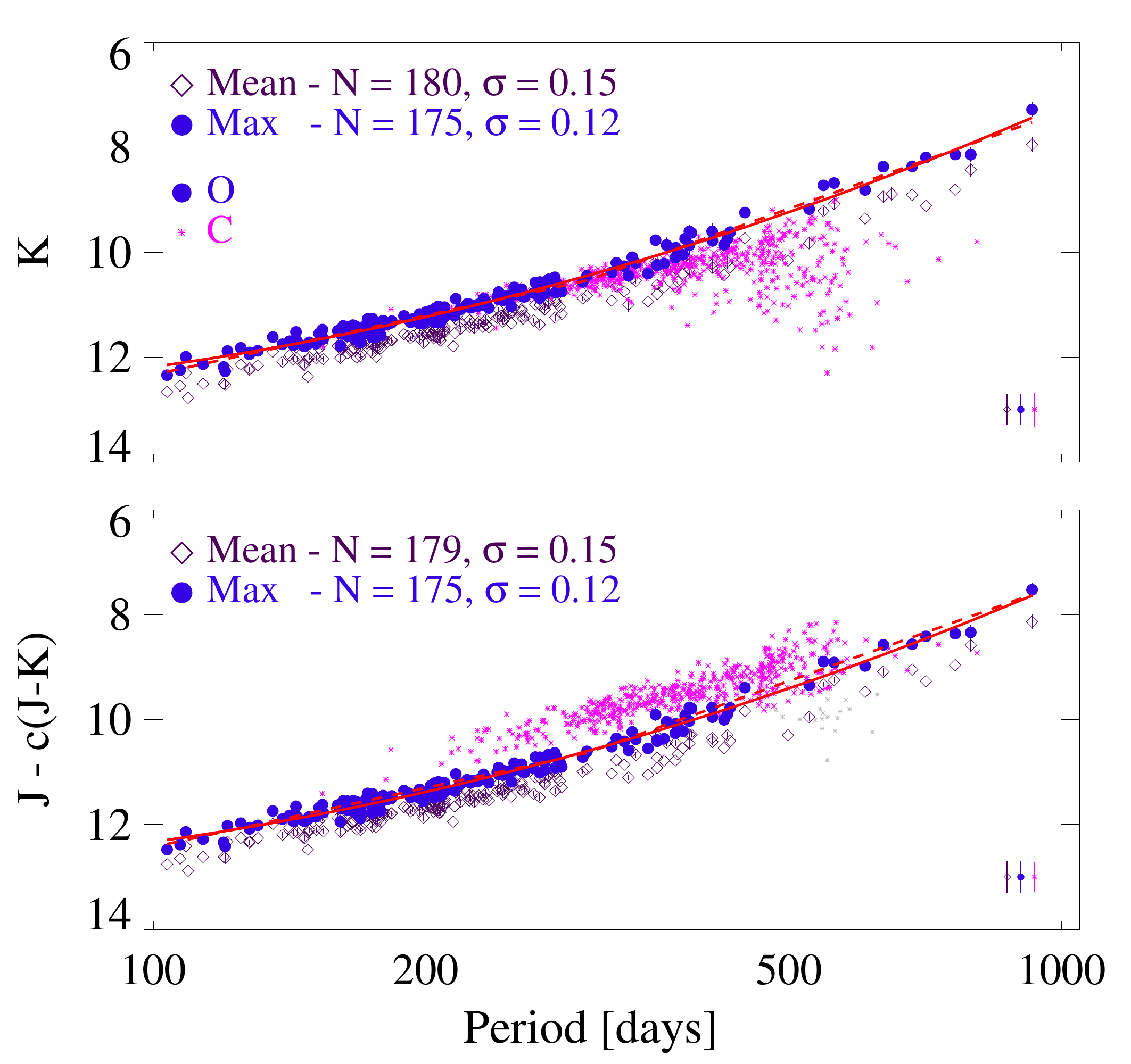} &
    \includegraphics[width=.45\textwidth]{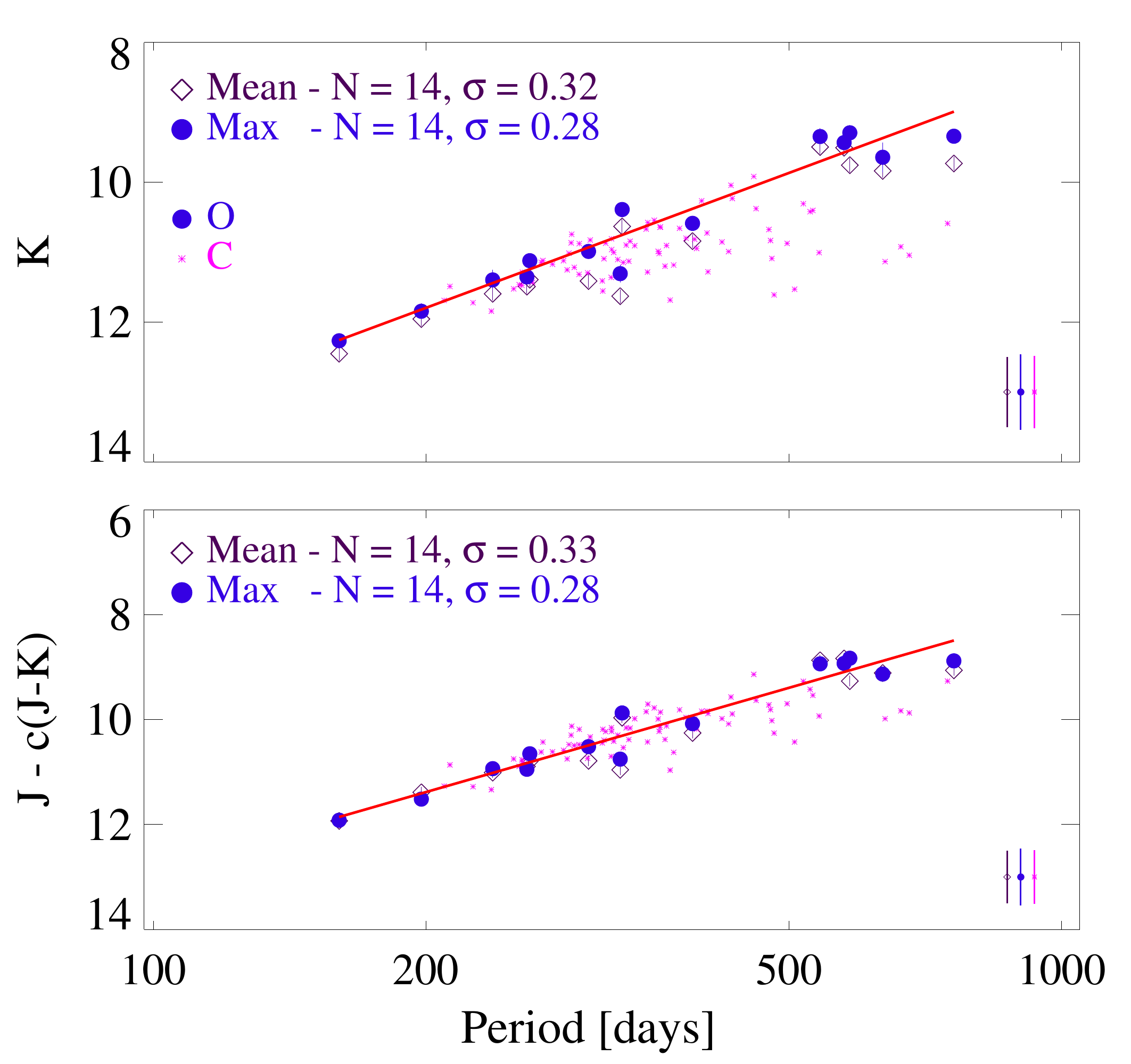} \\
  \end{tabular}
 \caption{{\it Left panel:} PL and PLC relations for Miras in the LMC using data from \citet{yuan2017b}. The solid/dashed line shows the best-fitting quadratic/non-linear regression model. 
For two-slope linear regression, a kink in the period is adopted at 300 days. Triangles and circles represent mean and max-light O-rich Mira PL/PLC relations while dots 
display the max-light PL/PLC relation for C-rich Miras. In the case of PLC relations, the color-coefficient ($c$) is different for O- and C-rich Miras (Table~\ref{tbl:nir_mc}).
{\it Right panel:} The same as the left panel but for SMC Miras using IRSF data \citep{ita2018}. The solid line represents the best-fitting linear regression model over the entire period range.
Representative $\pm 5\sigma$ errors are shown in each panel.}
\label{fig:nir_pl_plc}
\end{figure*}

Spectro-interferometry studies of O-rich Miras show that stellar radius in continuum light changes from max- to min-light \citep{thompson2002, wittkowski2016, wittkowski2018}. 
While the continuum and water radii are anticorrelated with the visual light curve, the Carbon monoxide (CO) bandhead diameter correlates well. The one-dimensional dynamic model 
atmospheres of Miras also predict an increase in the continuum radius between max and min-light, when the luminosity decreases, as well as an irregular variability in the outer 
molecular layers \citep[see,][]{ireland2011}. As the continuum radius decreases and becomes smaller at max-light, the effective temperature increases and dominates the visual 
light curve. Unstable molecules such as water vapor and TiO get destroyed at maximum effective temperature and luminosity. More stable CO molecules instead become more excited 
while going from min- to max-light through the pulsation cycle. The formation of different molecules may vary in the extended atmosphere from cycle to cycle, which 
contributes to the stellar luminosity and in turn to the variability between different pulsation cycles \citep{wittkowski2018}. 
At min-light, metallic oxides, such as TiO form sufficiently to extend 
the visual photosphere almost twice its nominal size, and decrease the visual light significantly \citep{reid2002}. Therefore, luminosity variations at max-light are significantly 
smaller than at mean or min-light possibly due to smaller contributions from  water vapor, TiO, and other unstable molecules at the warmest phase.

\begin{deluxetable*}{llrrrrrrrr}
\tablecaption{NIR PL and PLC relations for Miras in the Magellanic Clouds. \label{tbl:nir_mc}}
\tabletypesize{\footnotesize}
\tablewidth{0pt}
\tablehead{\colhead{MC} & \colhead{Band} & \colhead{O/C}& \colhead{$a$} & \colhead{$b$} & \colhead{$b_l$} & \colhead{$c$} & \colhead{$\sigma$} & \colhead{$N_i$}  & \colhead{$N_f$}}
\startdata
\cline{1-10}
\multicolumn{10}{c}{{Quadratic regression}}\\
\cline{1-10}
LMC          & $J$& O &      12.04$\pm$0.02      &	$-$3.83$\pm$0.09      &	$-$1.40$\pm$0.29      &                 ---      &       0.15&          180&          173\\
           & $J_M$& O &      11.62$\pm$0.02      &	$-$4.31$\pm$0.09      &	$-$1.58$\pm$0.31      &                 ---      &       0.17&          180&          178\\
       & $J,(J-K)$& O &      10.98$\pm$0.10      &	$-$4.31$\pm$0.09      &	$-$1.94$\pm$0.28      &       0.90$\pm$0.09      &       0.15&          180&          179\\
& $J_M,(J_M- K_M)$&O &      10.60$\pm$0.08      &	$-$4.82$\pm$0.08      &	$-$2.51$\pm$0.24      &       0.87$\pm$0.07      &       0.12&          180&          175\\
             & $K$& O &      10.86$\pm$0.01      &	$-$4.42$\pm$0.08      &	$-$2.06$\pm$0.27      &                 ---      &       0.15&          181&          180\\
           & $K_M$& O &      10.45$\pm$0.01      &	$-$4.87$\pm$0.07      &	$-$2.60$\pm$0.22      &                 ---      &       0.12&          181&          175\\
\cline{1-10}
\multicolumn{10}{c}{Linear regression}\\
\cline{1-10}
LMC          & $J$& O &      12.11$\pm$0.02      &	$-$3.15$\pm$0.10      &	$-$4.89$\pm$0.19      &                 ---      &       0.15&          180&          173\\
           & $J_M$& O &      11.68$\pm$0.02      &	$-$3.60$\pm$0.11      &	$-$5.31$\pm$0.21      &                 ---      &       0.16&          180&          176\\
       & $J,(J-K)$& O &      11.06$\pm$0.10      &	$-$3.43$\pm$0.10      &	$-$5.34$\pm$0.20      &       0.90$\pm$0.09      &       0.14&          180&          179\\
& $J_M,(J_M- K_M)$&O &      10.64$\pm$0.08      &	$-$3.76$\pm$0.08      &	$-$6.16$\pm$0.17      &       0.92$\pm$0.07      &       0.12&          180&          176\\
             & $K$& O &      10.94$\pm$0.02      &	$-$3.48$\pm$0.09      &	$-$5.53$\pm$0.18      &                 ---      &       0.14&          181&          179\\
           & $K_M$& O &      10.54$\pm$0.02      &	$-$3.76$\pm$0.08      &	$-$6.16$\pm$0.15      &                 ---      &       0.12&          181&          176\\
       & $J,(J-K)$& C &       9.67$\pm$0.05      &	$-$4.14$\pm$0.15      &                 ---      &       1.64$\pm$0.02      &       0.25&          488&          474\\
& $J_M,(J_M- K_M)$&C &       9.94$\pm$0.04      &	$-$4.25$\pm$0.14      &                 ---      &       1.35$\pm$0.02      &       0.25&          488&          470\\
\cline{1-10}
SMC          & $J$& O &      12.29$\pm$0.09      &	$-$4.34$\pm$0.44      &                 ---      &                 ---      &       0.35&           14&           14\\
           & $J_M$& O &      12.00$\pm$0.09      &	$-$4.45$\pm$0.41      &                 ---      &                 ---      &       0.34&           14&           14\\
       & $J,(J-K)$& O &      10.58$\pm$0.86      &	$-$4.75$\pm$0.44      &                 ---      &       1.55$\pm$0.78      &       0.33&           14&           14\\
& $J_M,(J_M- K_M)$&O &      10.50$\pm$0.54      &	$-$4.97$\pm$0.38      &                 ---      &       1.43$\pm$0.52      &       0.28&           14&           14\\
             & $K$& O &      11.19$\pm$0.08      &	$-$4.61$\pm$0.40      &                 ---      &                 ---      &       0.32&           14&           14\\
           & $K_M$& O &      10.94$\pm$0.07      &	$-$4.83$\pm$0.33      &                 ---      &                 ---      &       0.28&           14&           14\\
       & $J,(J-K)$& C &      10.28$\pm$0.15      &	$-$3.09$\pm$0.36      &                 ---      &       1.50$\pm$0.06      &       0.28&           76&           76\\
& $J_M,(J_M- K_M)$&C &      10.46$\pm$0.13      &	$-$3.15$\pm$0.39      &                 ---      &       1.36$\pm$0.07      &       0.30&           76&           76\\
\enddata
\tablecomments{The coefficients ($a, b, b_l, c$) are defined in Section~\ref{sec:data}. $\sigma$: dispersion (mag). $N_i$: initial number of sources. $N_f$: final number after rejecting extreme outliers.}
\end{deluxetable*}

\section{Miras at Infrared wavelengths} \label{sec:mira_nir}

Miras are known to exhibit tight PLRs at NIR wavelengths, specially in the $K$-band, as this band is least affected by atmospheric bands and the temperature fluctuations are 
significantly smaller \citep{feast1989, yuan2017b}. However, limited time-series data is available for LPVs at NIR wavelengths. The potential application
of max-light PLRs for distance scale studies is discussed in the following subsections.

\subsection{NIR PL and PLC relations at max-light}

Recently, \citet{yuan2017b} derived PLRs in the $JHK_s$ bands for Miras in the LMC based on the synoptic survey data from \citet{macri2015}. 
They generated NIR templates for OGLE Miras based on $I$-band light curves and derived robust mean, max, and min-light magnitudes from only three 
$JHK_s$ epochs. While the templates do not necessarily reflect the true shapes of the infrared light curves, they provide a reasonable approximation to the 
$JHK_s$ light curve structure with respect to that in the $I$-band. When templates are available, mean and max-light magnitudes are adopted from \citet{yuan2017b}.
If not, max-light magnitudes from the original light curves are estimated as discussed in Section~\ref{sec:data}. The left panel of Fig.~\ref{fig:nir_pl_plc} displays the 
mean and max-light PL and PLC relations for Miras in the LMC. At NIR wavelengths, both max- and mean-light PLRs show comparable scatter. The results of the regression
analysis are tabulated in Table~\ref{tbl:nir_mc}. Note that the mean-light NIR PLRs derived in this work are presented to compare with the max-light relations, and readers are referred
to \citet{yuan2017b} for calibrated PLRs. In the case of PLC relations, a significant color dependence is found only in the $J$-band for O-rich Miras \citep{feast1989}, and the color dependence
decreases at longer wavelengths.

While the LMC NIR data for Miras are limited in terms of temporal coverage, \citet{ita2018} provided NIR time series for a small region of the SMC spanning over a decade of observing epochs. Their
catalog consists of 14 O-rich and 76 C-rich Miras that have OGLE-like light curves in the $JHK_s$-bands. The right panel of Fig.~\ref{fig:nir_pl_plc} shows the PL and PLC relations for 
Miras in the SMC. The PLRs display evidence of a smaller dispersion at max-light than at mean-light despite the statistically small sample of Miras. 
The color coefficients of PLC relations for C-rich Miras are similar for both Magellanic Clouds and significant even at longer wavelengths.

\begin{figure}
\centering
  \begin{tabular}{@{}c@{}}
    \includegraphics[width=.45\textwidth]{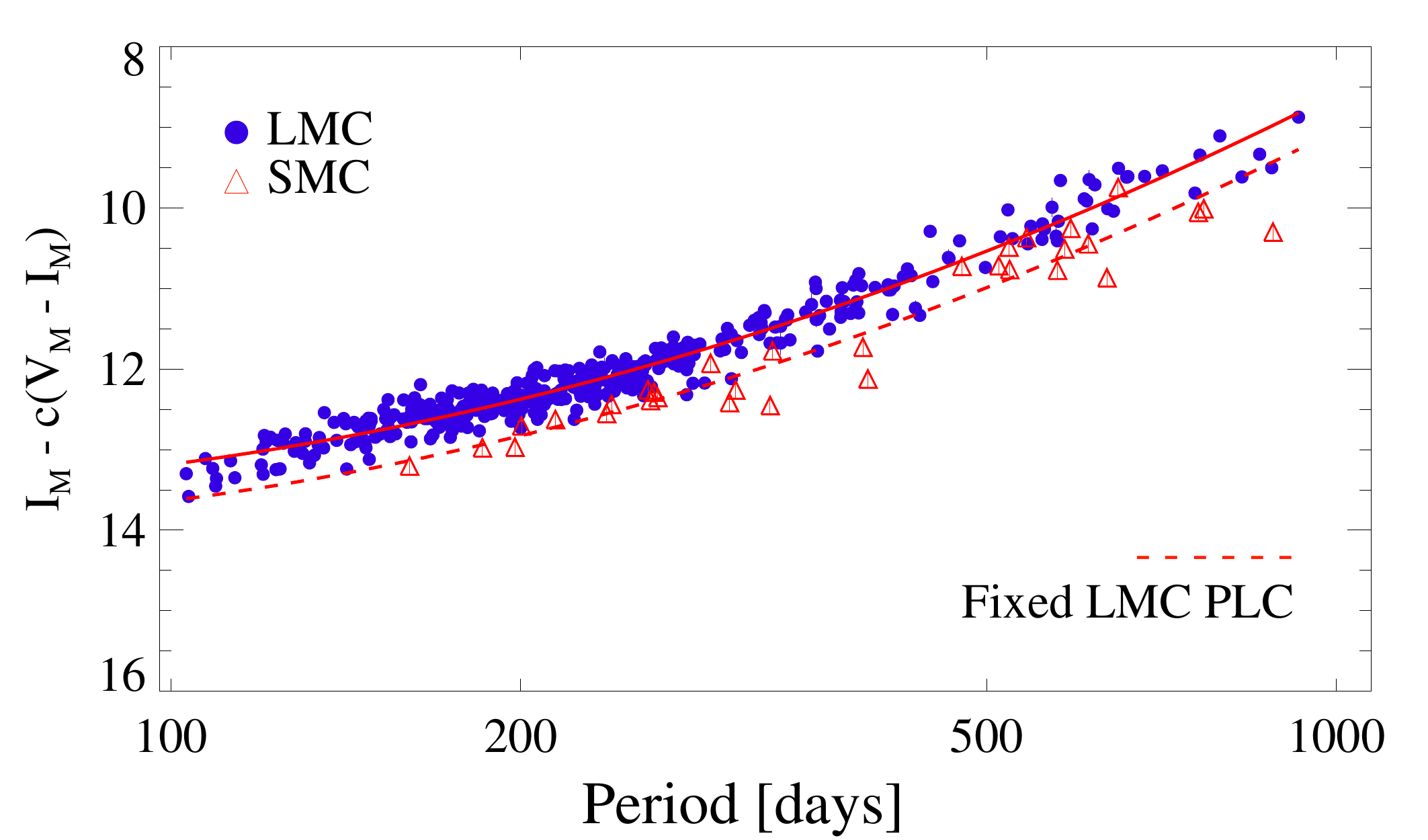} \\
    \includegraphics[width=.45\textwidth]{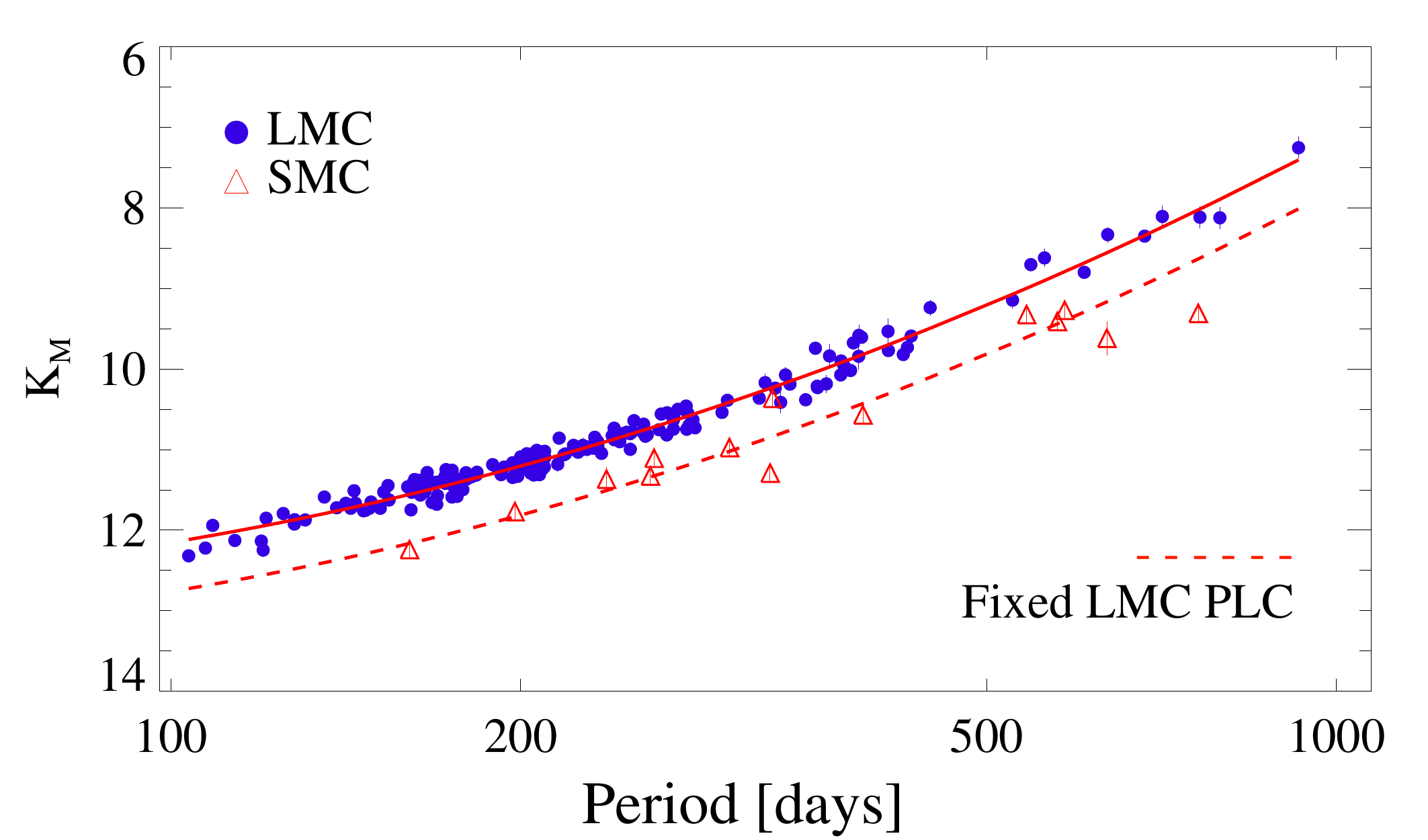} \\
  \end{tabular}
 \caption{Optical PLC and NIR PLR for O-rich Miras in the Magellanic Clouds. The dashed line indicates the fixed-LMC PL/PLC used to estimate the relative distance modulus between the
Magellanic Clouds.}
 \label{fig:mu_mc}
\end{figure}

\subsection{Distance between the Magellanic Clouds}

O-rich Mira PLRs at NIR wavelengths can be used as potential distance indicators. While the long-period Miras are affected by the circumstellar dust extinction,
the extinction and metallicity effects are small at NIR wavelengths. Assuming that the effects of circumstellar dust are similar at both mean and max-light, we 
used O-rich Mira PLRs in the LMC and SMC to estimate the relative distance between the Magellanic Clouds. 
The top panel of Fig.~\ref{fig:mu_mc} presents the optical PLC relations and the bottom panel displays NIR $K_s$-band PLRs at max-light for O-rich Miras in the Magellanic Clouds. 
The zero-point offset between the Clouds is estimated with respect to fixed LMC relations. Table~\ref{tbl:mu_mc} lists the results for 
the relative distance modulus between the LMC and the SMC estimated using both mean and max-light relations. The extinction-corrected
distance estimates based on max-light relations exhibit smaller statistical uncertainties by virtue of the smaller dispersion in max-light relations. A mean relative distance modulus, 
$\Delta \mu = 0.48\pm0.08$~mag, is obtained between the Magellanic Clouds using O-rich Miras. The distance estimate using max-light PLRs are consistent with the 
relative distance moduli based on classical Cepheids, Miras, and other distance indicators \citep{matsunaga2012, degrijs2014, degrijs2015, bhardwaj2016a}.

\begin{deluxetable}{lrrrr}
\tablecaption{Relative distance moduli for the Magellanic Clouds. \label{tbl:mu_mc}}
\tabletypesize{\footnotesize}
\tablewidth{0pt}
\tablehead{\colhead{Band} & \colhead{ZP} & \colhead{$\Delta_\mu$}& \colhead{ZP} & \colhead{$\Delta_\mu$}}
\startdata
PL/PLC &\multicolumn{2}{c}{{Mean-light}} & \multicolumn{2}{c}{{Max-light}}\\
\cline{1-5}
              $I,(V-I)$&       12.07$\pm$0.09&       0.35$\pm$0.12&      12.12$\pm$0.07&       0.45$\pm$0.08\\
                    $J$&      12.37$\pm$0.09&       0.33$\pm$0.11&      12.08$\pm$0.07&       0.46$\pm$0.08\\
                    $K$&      11.27$\pm$0.08&       0.41$\pm$0.09&      10.98$\pm$0.06&       0.53$\pm$0.07\\
\cline{1-5}
\enddata
\end{deluxetable}

\subsection{Distance to the Galactic Center}

\begin{figure}
\epsscale{1.2}
\plotone{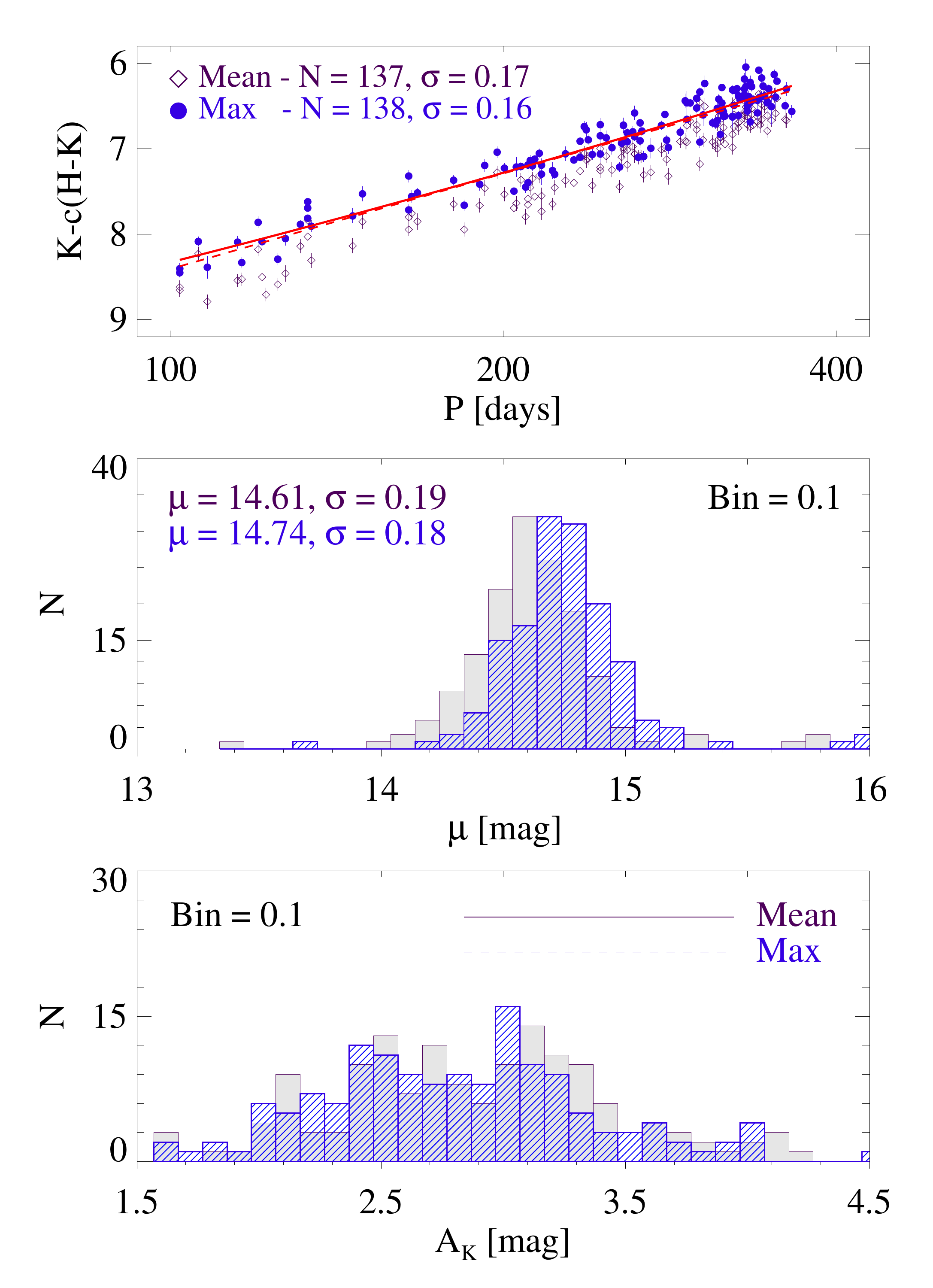}
\caption{{\it Top panel:} PLC relation for Miras in the Galactic Center from \citet{matsunaga2009}. The solid/dashed line shows the best-fitting quadratic/linear regression model. 
Diamonds and circles represent mean and max-light O-rich Mira PL/PLC relations. {\it Middle panel:} Histograms of estimated distance moduli of Miras in the 
Galactic Center using calibrated mean- and max-light PLRs in the LMC. {\it Bottom panel:} Histograms of estimated $K$-band extinction using a similar approach. \label{fig:gc_lpv}}
\end{figure}

\citet{matsunaga2009} carried out a NIR survey of LPVs towards the Galactic Center and identified 1364 variables, including several Mira candidates. Due to significantly high extinction, 
$J$-band light curves are not available for several variables while $K$-band magnitudes approach the saturation limit for the brightest LPVs. We selected a subsample of Miras with
the following criteria - $P\le$~350 days, max-light magnitudes $\ge 9$~mag in $K$, amplitudes greater than $0.5$~mag in $K$. These criteria led to a sample of 148 Miras with both $H$- and $K$-band measurements, which is used in the subsequent analysis. We assume that most of these Miras are O-rich as C-rich Miras are rare towards the central region of the Galaxy \citep{matsunaga2017}.

Fig.~\ref{fig:gc_lpv} displays the PLC relation for Miras towards the Galactic Center. The scatter is comparable in both mean- and max-light relations. \citet{matsunaga2009} 
adopted total-to-selective absorption ratios (e.g., $A_H/A_K$) given by the reddening law of \citet{nishiyama2006} to estimate distance and extinction, simultaneously.
Similar to their analysis, we used equations 8 and 9 from \citet{matsunaga2009} to estimate the distance and $K$-band extinction for 
Miras with both max- and mean-light PLRs. The calibrator NIR data for Miras in the LMC are taken from \citet{yuan2017b} and the PLRs are restricted to Miras 
with $P\le$~350 days. Mira PLRs are calibrated using the most-precise $\sim 1\%$ distance to the LMC \citep{piet2019}. 
The middle panel of Fig.~\ref{fig:gc_lpv} shows the histogram of distances to individual
Miras towards the Galactic Center. Our distance estimates using mean-light are similar to that of \citet{matsunaga2009} and the peak of the distribution is consistent with the 
$\sim 0.3\%$ geometric distance ($\sim 8.18$~kpc) to the Galactic center \citep{gravity2019}, and with measurements based on RR Lyrae and Type II 
Cepheids \citep{dekany2013, bhardwaj2017b}. However, the peak of the distribution 
using max-light PLRs shows an offset towards larger distances but is consistent within the uncertainties. The bottom panel of Fig.~\ref{fig:gc_lpv} shows the $K$-band extinction estimated using
mean and max-light PLRs. The typical values and the range of $A_K$ are consistent between the two approaches and similar to the results of \citet{matsunaga2009}.

\section{Conclusions and Discussion} \label{sec:discuss}

The pulsation properties of Mira variables in the Magellanic Clouds are investigated as a function of the pulsation phase, in particular at maximum-light.  The max- and mean-light
Period-Luminosity and Period-Luminosity-Color relations for Mira variables are compared for the first-time at multiple wavelengths. 
The max-light PL and PLC relations exhibit significantly smaller (up to 30\%)
dispersion than at mean-light at optical wavelengths. With limited near-infrared data, the scatter in max-light relations is comparable to mean-light relations at longer wavelengths. 
The typical variations in the luminosity (magnitudes) and temperatures (colors) for Miras with similar periods are smaller at their brightest phase. Further, the max-light magnitudes 
for each Mira over different pulsation cycles are also more stable than the min-light magnitudes. The spectral features of O-rich Miras at max-light are dominated by strong Balmer 
line emission and TiO molecular bands. The strength of TiO bands is weaker at redder wavelengths and increases from max- to min-light. 
The stability of max-light magnitudes for individual Miras could be due to lower sensitivity to molecular bands in their warmest phase. On the other hand, the stability 
of max-light for a group of Miras with similar periods may also indicate that Miras achieve similar maximum luminosity corresponding to their primary period in each pulsation cycle.

Mira variables offer exciting prospects for the extragalactic distance scale work in the Large Synoptic Survey Telescope (LSST) era. The identification and classification of Miras are complicated
due to their long-periods that require monitoring for several years. Although LSST will provide time-domain multi-color photometry, optical PLRs for
Miras exhibit large scatter. The high mass-loss rate in these luminous AGB stars forms thick dust shells. Consequently, Miras suffer from significant circumstellar extinction at shorter wavelengths. 
Therefore, infrared observations are essential to reduce these effects and to derive robust PLRs. LSST will provide a catalog of Miras in external galaxies, 
and their infrared luminosities from the upcoming ground-based large-aperture telescopes and space-based James Webb Space Telescope (JWST) will be ideal for future distance scale studies. 
While the traditional approach of dealing with mean-light properties will require at least a few epochs of infrared observations, the stability of max-light can provide 
an easier and simpler alternative. Additionally, max-light observations will also permit a probe of external galaxies at much larger distances, and therefore, an assessment of their stability 
and reliability in other stellar systems warrant further investigation.

\acknowledgements
We thank the anonymous referee for the quick and constructive referee report that helped improve the paper. We also thank Evelin B\'anyai, Attila B\'odi, and Tim Bedding for sending 
the M-giants and Mira light curves from the {\it Kepler} mission, Wenlong Yuan and Lucas Macri for reading an earlier version of the manuscript, and M. Wittkowski for helpful discussions.
AB acknowledges research grant $\#11850410434$ awarded by the National Natural Science Foundation of China through the Research Fund for International Young Scientists,
and a China Post-doctoral General Grant. SMK and HPS acknowledge the support from the Indo-US Science and Technology Forum, New Delhi, India. 
NM is grateful to Grant-in-Aid (KAKENHI, No. 18H01248) from the Japan Society for the Promotion of Science (JSPS).
CCN and JYO thank the support from the Ministry of Science and Technology (Taiwan) under grant 107-2119-M-008-014-MY2.
This research was supported by the Munich Institute for Astro- and Particle Physics (MIAPP) of the DFG cluster of excellence ``Origin and Structure of the Universe''. 

\bibliographystyle{aasjournal}
\bibliography{/home/anupam/work/manuscripts/mybib_final.bib}
\end{document}